\documentclass{article}
\usepackage{url}

\usepackage{graphicx}
\usepackage{algorithm}
\usepackage{algorithmic}
\usepackage{amsmath,amssymb,amsfonts}
\usepackage{amsthm}
\usepackage{array}
\usepackage{bm}
\usepackage{cite}
\usepackage[varg]{txfonts}
\usepackage{wrapfig}
\usepackage{color}
\usepackage{authblk}
\usepackage[margin=25mm]{geometry}

\newcommand{\abs}[1]{\left\lvert #1 \right\rvert}

\DeclareMathOperator*{\argmax}{arg\,max}

\newcommand{\ReviewerA}{black}
\newcommand{\ReviewerB}{black}
\newcommand{\ReviewerAA}{black}
\newcommand{\ReviewerBB}{black}
\newcommand*{\reviewerA}[1]{\textcolor{black}{#1}}%English Blue
\newcommand*{\reviewerB}[1]{\textcolor{black}{#1}}%Reviewer-1 (3rd
\theoremstyle{definition}
\newtheorem{dfn}{Definition}[section]

\newtheorem{thm}[dfn]{Theorem}

\newtheorem{rem}[dfn]{Remark}

\providecommand{\keywords}[1]{\textbf{\textit{Index terms---}} #1}

\begin{document}

\title{Nondominated-Solution-based\\ Multi-objective Greedy Sensor Selection\\ for Optimal Design of Experiments}

% \author{
% Kumi Nakai, \IEEEmembership{nonmember, IEEE}, 
% Yasuo Sasaki, \IEEEmembership{nonmember, IEEE}, 
% Takayuki Nagata, \IEEEmembership{nonmember, IEEE},
% Keigo Yamada, \IEEEmembership{nonmember, IEEE},  
% Yuji Saito, \IEEEmembership{nonmember, IEEE}, 
% Taku Nonomura, \IEEEmembership{nonmember, IEEE}
% %\thanks{This work was supported in part by JST, and Japan. }
% \thanks{
% K. Nakai is with National Institute of Advanced Industrial Science and Technology, Tsukuba, 305-8564, Japan and with Tohoku University, Sendai 980-8579, Japan (e-mail: kumi.nakai@aist.go.jp), 
% Y. Sasaki is with Tohoku University, Sendai 980-8579, Japan (e-mail: yasuo.sasaki.e4@tohoku.ac.jp), 
% T. Nagata is with Tohoku University, Sendai 980-8579, Japan (e-mail: nagata@tohoku.ac.jp), 
% K. Yamada is with Tohoku University, Sendai 980-8579, Japan (e-mail: keigo.yamada.t5@dc.tohoku.ac.jp), 
% Y. Saito is with Tohoku University, Sendai 980-8579, Japan (e-mail: yuji.saito@tohoku.ac.jp), 
% T. Nonomura is with Tohoku University, Sendai 980-8579, Japan (e-mail: nonomura@tohoku.ac.jp).}
% }

\author[1,2]{Kumi Nakai\footnote{Corresponding to kumi.nakai@aist.go.jp}}
\author[2]{Yasuo Sasaki}
\author[2]{Takayuki Nagata}
\author[2]{Keigo Yamada}
\author[2]{Yuji Saito}
\author[2]{Taku Nonomura}
\affil[1]{National Institute of Advanced Industrial Science and Technology}
\affil[2]{Graduate School of Engineering, Tohoku University}

\markboth{Journal of \LaTeX\ Class Files, Vol. xx, No. xx, xxxx 202x}
{Shell \MakeLowercase{\textit{et al.}}: Bare Demo of IEEEtran.cls for IEEE Journals}
\maketitle
% \linenumbers
\begin{abstract}
In this study, a nondominated-solution-based multi-objective greedy method is proposed and applied to a sensor selection problem based on the multiple indices of the optimal design of experiments. The proposed method simultaneously considers multiple set functions and applies the idea of Pareto ranking for the selection of sets. Specifically, a new index  is iteratively added to the nondominated solutions of sets, and the multi-objective functions are evaluated for new sets. The nondominated solutions are selected from the examined solutions, and the next sets are then considered. With this procedure, the multi-objective optimization of multiple set functions can be conducted with reasonable computational costs.
%added 2022 06 09 takunonomura
\reviewerA{This paper defines a new class of greedy algorithms which includes the proposed nondominated-solution-based multi-objective greedy algorithm and the group greedy algorithm, and the characteristics of those algorithms are theoretically discussed.} 
%end added 2022 06 09 takunonomura
Then, the proposed method is applied to the sensor selection problem and its performance is evaluated. The results of the test case show that the proposed method not only gives the Pareto-optimal front of the multi-objective optimization problem but also produces sets of sensors in terms of D-, A-, and E-optimality, that are superior to the sets selected by pure greedy methods that consider only a single objective function. 
\end{abstract}

% \begin{IEEEkeywords}
% Greedy algorithm, Multi-objective optimization, Nondominated solution, Pareto front, Optimal experimental design, Data-driven sensor selection
% \end{IEEEkeywords}
\renewcommand{\thefootnote}{}
\footnote[0]{This paper has been accepted for publication in IEEE TRANSACTIONS ON SIGNAL PROCESSING. This manuscript version is made available under the CC-BY-NC-ND 4.0 license (https://creativecommons.org/licenses/by-nc-nd/4.0/)}

% \IEEEpeerreviewmaketitle

\section{Introduction}

% \IEEEPARstart{T}{he} 
The sensor selection problem has received considerable attention in the area of distributed parameter systems such as fluid dynamics. The sensor selection problem can be interpreted as a combinatorial problem. To implement rigid optimization for such a problem is computationally expensive. For instance, if a brute-force search to select 25 sensors from 100 sensor candidates were to be conducted,  $\tiny{ \left( \begin{array}{c}100\\ 25 \end{array}\right)}\approx 2.4\times 10^{23}$ combinations would need to be investigated. Given the massive number of possibilities even in such a small version of the problem, a direct search approach is clearly untenable and computationally efficient ways of selecting sensors need to be found. To date, several computational methods have been proposed for sensor selection, especially for linear observation problems. 

% Optimal design of experiment
The quality of a set of sensors can be evaluated using the optimality criteria proposed in the optimal design of experiments\cite{atkinson2007optimum}. \color{\ReviewerB}These \color{black} criteria are defined using the Fisher information matrix (FIM), which corresponds to the inverse of the error covariance matrix of estimation using sensors. The most commonly used criterion is D-optimality, where the determinant of the FIM is maximized. This criterion results in the minimization of the volume of the confidence ellipsoid of the regression estimates. Two other criteria also have a statistical interpretation in terms of the FIM: A-optimality, in which the trace of the inverse of the FIM is minimized, and E-optimality, where the minimum eigenvalue of the FIM is maximized. When considering the optimal set of sensors, D-, A-, and E-optimality are often employed, and the sensors are selected so that the index of the optimality criterion can be optimized. 

% Introduction of sensor selection method
Previously, the optimization of a set of sensors based on the optimal design of experiments has been limited due to lack of numerical tools. More recently, however, such optimizations can be conducted using newly developed numerical tools, especially focusing on the D-optimality criterion. 
Joshi and Boyd defined an objective function and proposed a convex relaxation of the problem, including a semi-definite programming method\cite{joshi2009sensor}. The convex relaxation method is then able to produce a global optimal solution to the relaxed problem. The drawback of the method is the computational complexity of $\mathcal{O}\left(n^3\right)$, where $n$ denotes the number of sensor candidates. The algorithm was recently improved \cite{nonomura2021randomized} with the development of a randomized algorithm\cite{gower2019rsn} for the acceleration. Convex optimization methods based on the proximal splitting algorithm have also been proposed, such as the alternating direction method of multipliers\cite{lin2013design,fardad2011sparsity,dhingra2014admm,zare2018optimala}. Although a sensor selection method based on the proximal splitting algorithm for the A-optimality criterion that significantly reduces the computational cost has been proposed\cite{nagata2021data,nagata2022data}, the computational cost is still large when applied to a many-degrees-of-freedom problem such as in the data-driven sensor selection of fluid dynamics. Alternatively, sensor selection methods based on greedy algorithm that are faster than the convex relaxation methods have been proposed\cite{manohar2018data,saito2021determinant,saito2020data,nakai2021effect}. 

% Greedy method
Manohar et al.\cite{manohar2018data} applied and extended the QR-based discrete empirical interpolation method\cite{chaturantabut2010nonlinear,drmac2016new}, which is one of the greedy methods, to the sensor selection problem. Saito et al.\cite{saito2021determinant} showed mathematically that the method proposed by Manohar et al.\cite{manohar2018data} is equivalent to the greedy method for the set function based on D-optimality when the number of sensors is less than the number of latent variables. They also proposed a greedy method based on D-optimality that is efficient regardless of the number of sensors. These greedy methods are able to find a solution that is within $(1-1/e)$ of the optimal solution when the objective function is monotone submodular, and they outperform convex relaxation methods when the size of the problem  increases\cite{ranieri2014near,jiang2016sensor,shamaiah2010greedy}. Due to their fast calculation and reasonable performance, such greedy methods are more appealing for sensor selection. They have been extended for various purposes\cite{liu2016sensor,clark2018greedy,peherstorfer2020stability,clark2020multi,clark2020sensor,manohar2021optimal,yamada2021fast,saito2021data-driven,yamada2022greedy,nagata2022randomized,li2022retrieval} and implemented to several applications\cite{kanda2021feasibility,kaneko2021data,inoue2021data-driven,inoba2022optimization,kanda2022proof,tiwari2022simultaneous, inoue2022data,li2021data,fukami2021global,callaham2019robust,turko2022information,yeo2022effcient,nakai2022observation,nagata2022seismic}. 
Nakai et al.\cite{nakai2021effect} evaluated the effects of objective set functions based on D-, A-, and E-optimality on optimization results using the greedy methods. They showed that especially the E-optimality-based greedy method does not work better than the D- and A-optimality-based greedy methods even in terms of E-optimality. They also reported that the A-optimality-based greedy method is better than the E-optimality-based greedy method even for maximizing the minimum eigenvalue when using a pure greedy method.
This illustrates the advantage of using a greedy method based on an optimality that is different from the optimality of interest. 
In addition, Nakai et al.\cite{nakai2021effect} reported that the optimality best suited to sensor selection depends on the characteristics of the dataset and the situation. %D- and A-optimality are important in the ideal case where the statistics for the systems are well known; on the other hand, E-optimality is critical in terms of practicality. This suggests that selecting the appropriate objective function very much depends on the situation. 

The other aspects of the outcomes of the series of studies by Saito et al.\cite{saito2021determinant} and Nakai et al.\cite{nakai2021effect} show that the heuristic high performance sensor selection methods such as the QR method are indeed formulated in the conventional forms of greedy methods of a set function. Therefore, improvement or relaxation in greedy method algorithms can lead to better selection of the sensor sets.  

% group greedy
A group greedy method recently has been proposed as a way to improve the greedy methods for set functions\cite{jiang2019group} and applied it to the sensor selection problem. The typical greedy method iteratively finds the one new index that produces the greatest improvement in the objective value of interest and adds it to the current best set. Such a step-by-step strategy can drastically reduce the computational cost but may miss some better sets of indices. In contrast, the group greedy method proposed by Jiang et al. \cite{jiang2019group} iteratively reserves not only the one best set but also retains suboptimal sets. Adding a new index to a suboptimal set from the previous step may outperform adding an index to the best set. The group greedy method were reported to outperform the typical pure greedy method in  sensor selection problems because of the extended search space in the group greedy strategy\cite{jiang2019group}. 

\begin{figure}[hb]
    \centering
    \includegraphics[width=3.4in]{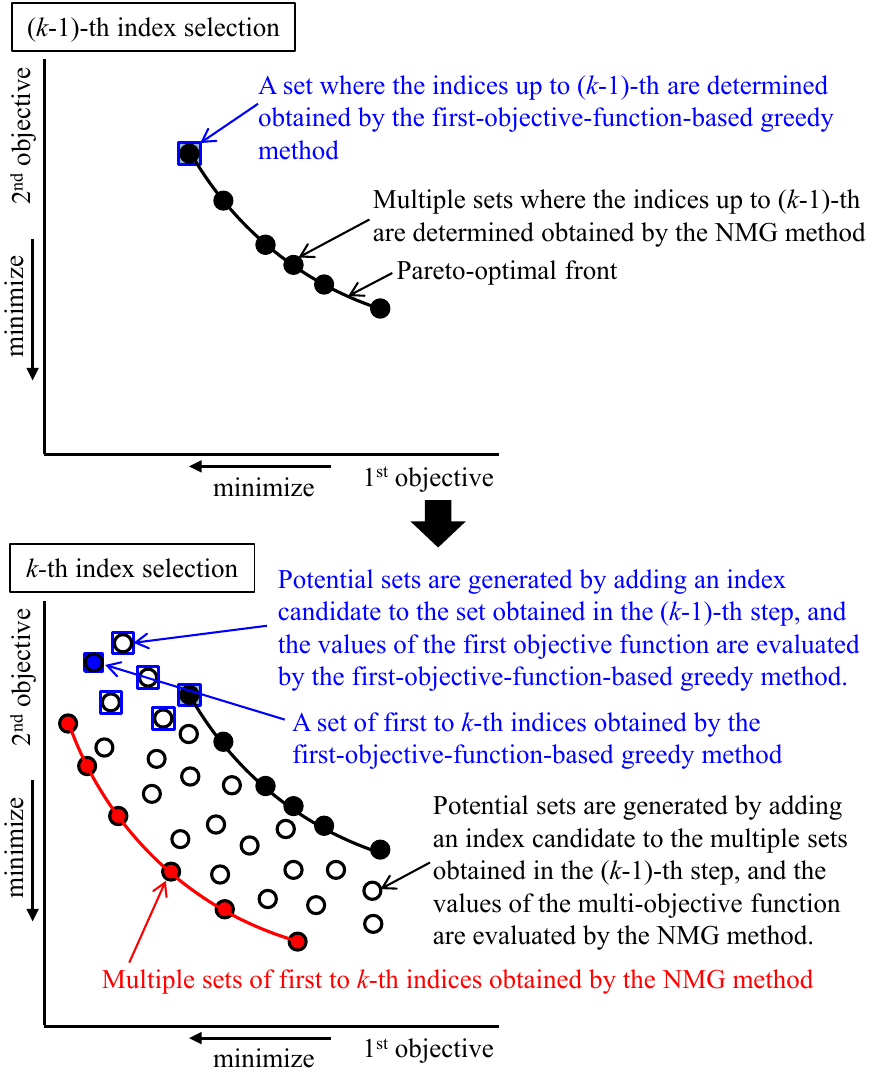}
    \caption {\color{\ReviewerAA}Graphical image of index selection using nondominated-solution-based multi-objective greedy method.\color{black}}
    \label{fig:concept}
\end{figure}

% Proposal of multi-objective greedy
% 手法の説明
Based on these previous studies, it would seem that a greedy method that produces nondominated solutions by solving a multi-objective optimization problem (MOP) could be expected to improve the performance of greedy methods. 
The multi-objective genetic algorithm (MOGA) is recognized as a powerful method for solving MOPs. A number of MOGAs have been developed\cite{horn1994niched,srinivas1994multiobjective,deb2002fast}, and the application of MOGA has been intently studied in the engineering field, e.g., the optimization of the airfoil for aircraft and turbines\cite{obayashi2000multiobjective,shimoyama2011multi-objective,liu2014optimization,tatsukawa2016optimisation,jeong2018optimization,wei2020research}.
%A graphical description of sensor selection using our proposed nondominated-solution-based multi-objective greedy method is shown in Fig.~\ref{fig:concept}. With the multi-objective greedy method, we iteratively add a new sensor to a set of sensors obtained in the previous step by simultaneously evaluating D-, A-, and E-optimality. 
For MOPs, there normally exists a set of trade-off solutions rather than one optimal solution. Thus, algorithms for solving MOPs typically produce nondominated solutions (Pareto-optimal solutions), which, belonging to the same rank (the Pareto-optimal front), are viewed to be equally important, using the notion of Pareto dominance. The nondominated-solution-based multi-objective greedy (NMG) method proposed in the present study is a multi-objective optimization method for the problem formulated by multiple set functions inspired by MOGA. A graphical description of our proposed method is shown in Fig.~\ref{fig:concept}. With the multi-objective greedy method, we iteratively add a new index to a set obtained in the previous step by simultaneously evaluating multiple objective set functions. The multi-objective greedy method iteratively finds and reserves sets corresponding to the nondominated solutions in a multi-objective-function space in each step of index selection.

% まずgruop-greedyと同様の効果
When applying it to the sensor selection problem, the NMG method is expected to improve the performance of solutions compared with the typical pure greedy method, which finds one new sensor in each single-sensor subproblem. This is due to the mechanism of increasing the search space, the effectiveness of which has been demonstrated by the above-mentioned group greedy strategy\cite{jiang2019group}.
% さらにgroup greedyを上回る多目的の効果
Furthermore, the method potentially provides better sets of sensors than the pure and group greedy methods based on a single objective function. The pure greedy method that considers only the optimality of interest may miss some better sets of sensors, as noted above\cite{nakai2021effect}, as even the group greedy method reserves multiple sets of sensors only in terms of the optimality of interest. The NMG method overcomes this limitation by considering not only the optimality of interest but also other optimality. It also enables the user to choose the appropriate set of sensors from the nondominated solutions based on the optimality of interest. 

% Objective
In this paper, we propose the NMG method for the multiple set functions and applied it to the sensor selection problem. First, the algorithm and the theoretical characteristics of the proposed NMG method are provided. Then, the proposed method is applied to the sensor selection problem. The proposed method applied to the sensor problem simultaneously evaluates D-, A-, and E-optimality for the potential sensors in the next step and determines the sets of sensors based on the nondominated solutions in the multi-objective-function space. The improvement in the indices of D-, A-, and E-optimality by the NMG method compared with those obtained by the pure and group greedy methods are presented.

The remainder of the paper is organized as follows: In Section~\ref{sec:probandalg}, the objective functions used in the present study, as well as the algorithm of the NMG method are explained. \reviewerA{In Section~\ref{sec:theory}, the class of relaxed greedy methods which includes the NMG method is defined and the characteristics of the method is theoretically discussed.} In Section~\ref{sec:res}, test results for the proposed method are discussed.  Section~\ref{sec:con} concludes the paper. 

The sensor optimization method can also be used for the actuator optimization problem, which is a dual problem with the sensor selection problem. The idea proposed in the present study can improve the performance of various optimizations using greedy methods, not just sensor/actuator optimization.

\section{Problem and Algorithm}
\label{sec:probandalg}
The sensor selection problem and the applicable objective functions based on the optimal design of experiments are described below, together with the multi-objective greedy algorithm using nondominated sorting. 

\subsection{Sensor Selection Problems}
We define the sensor selection problem as follows:
\begin{align}
    \mathbf{y}=\mathbf{H}\mathbf{U}\mathbf{z}=\mathbf{C}\mathbf{z},
\end{align}
where $\mathbf{y} \in \mathbb{R}^{p} $ is the observation vector, $\mathbf{H} \in \mathbb{R}^{p \times n}$ is the sensor location matrix, $\mathbf{U} \in \mathbb{R}^{n \times r}$ is the sensor candidate matrix, $\mathbf{z} \in \mathbb{R}^{r}$ is the latent variable vector, and $\mathbf{C} \in \mathbb{R}^{p \times r}$ is the measurement matrix ($\mathbf{C}=\mathbf{HU}$). Here, $p$, $n$, and $r$ are the number of sensors to be selected, the total number of sensor candidates, and the number of latent variables, respectively. The sensor location matrix $\mathbf{H}$ has unity at the sensor location and zeros at the other locations for each row. As described in\cite{nakai2021effect}, the objective function for sensor selection problems can be defined using the FIM based on the optimal design of experiments. As in the previous study, the D-, A-, and E-optimality criteria are considered here. 
The objective function based on D-optimality can be expressed as follows:
\begin{align}
    &\mathrm{maximize}\,\,f_{\mathrm{D}} \nonumber \\
    &f_{\mathrm{D}}\,=\,\left\{\begin{array}{cc}
        \mathrm{det}\,\left(\mathbf{C}\mathbf{C}^{\top}\right), & p\le r, \\
        \mathrm{det}\,\left(\mathbf{C}^{\top}\mathbf{C}\right), & p>r.
    \end{array}\right.\label{eq:obj_det}
\end{align}
The objective function based on A-optimality is as follows:
\begin{align}
    &\mathrm{minimize}\,\,f_{\mathrm{A}} \nonumber \\
    &f_{\mathrm{A}}\,=\,\left\{\begin{array}{cc}
        \mathrm{tr}\,\left[\left(\mathbf{C}\mathbf{C}^{\top}\right)^{-1}\right], & p\le r, \\
        \mathrm{tr}\,\left[\left(\mathbf{C}^{\top}\mathbf{C}\right)^{-1}\right], & p>r.
    \end{array}\right.\label{eq:obj_tr}
\end{align}
Finally, the objective function based on E-optimality can be expressed as:
\begin{align}
    &\mathrm{maximize}\,\,f_{\mathrm{E}} \nonumber \\
    &f_{\mathrm{E}}\,=\,\left\{\begin{array}{cc}
        \lambda_{\mathrm{min}}\,\left(\mathbf{C}\mathbf{C}^{\top}\right), & p\le r, \\
        \lambda_{\mathrm{min}}\,\left(\mathbf{C}^{\top}\mathbf{C}\right), & p>r.
    \end{array}\right.\label{eq:obj_eig}
\end{align}
Note that simplified formulations are employed using the greedy method as discussed in the previous studies\cite{saito2021determinant,nakai2021effect}.
Here, D-, A-, and E-optimality functions could be formulated in the form of a set function of indices of sensors selected, and the pure greedy method for a set function can be straightforwardly applied to each objective function.

\subsection{Nondominated-solution-based Multi-objective Greedy Algorithm}
In the present study, we are proposing an NMG method  for solving MOPs for multiple set functions, and applying it to a sensor selection problem based on simultaneous optimal experimental design, i.e., D-, A-, and E-optimality. 

\color{\ReviewerA}Algorithm \ref{alg:nmg} shows the procedure of the NMG method for multiple set functions.
In the conventional greedy method for a set function, selection of only the \color{\ReviewerBB}$k$--th \color{black} index is carried out in the \color{\ReviewerBB}$k$--th \color{black}step for a set where the indices up to \color{\ReviewerBB}($k$-1)--th \color{black} are already determined. In the NMG method, selection of the \color{\ReviewerBB}$k$--th \color{black} indices is carried out for the $L_{\rm{\max}}$ sets where the indices up to \color{\ReviewerBB}($k$-1)--th \color{black} are already determined. In the first step of the index selection by the NMG method, the values of multi-objective functions are calculated for all the $n$ index candidates. Then, the top $L_{\rm{\max}}$ sets are selected according to the order of the Pareto rank in MOP using a nondominated sorting. The nondominated sorting and candidate selection will be discussed in detail in this subsection. In the second step, potential sets that are generated by adding an index candidate to each of the $L_{\rm{\max}}$ sets obtained in the first step are considered, and the values of the multi-objective functions of all the potential sets are evaluated. Here, the duplication of the potential sets is removed. Then, the top $L_{\rm{\max}}$ sets are selected using a nondominated sorting. The procedure above is repeatedly conducted until a particular constraint is satisfied, such as when the number of selected indices $k$ reaches a predefined number $p$. Finally, we obtain $L_{\rm{\max}}$ sets of indices in which first to \color{\ReviewerBB}$p$--th \color{black} indices are determined. \color{black} 
%% Introductionから移動：
%This indicates that a greedy method based on a target optimality may not be able to obtain an optimal sensor set for the target optimality, depending on the conditions. Greedy methods based on optimality different from the target one may work better to select an optimal sensor for the target one.  
%The reconstruction error using sensors selected by the greedy methods based on different optimal criteria has been compared. As a result, the indices of D- and A-optimality are important in the ideal case where the statistics for the systems are well known, and on the other hand, the index of E-optimality is critical in the practical case. 
%%

As discussed earlier, the typical pure greedy method reserves a set of indices in each step. It finds the best index by considering index candidates only for the set obtained in the preceding step and evaluating a single objective function. 
% a particular constraint is satisfied, such as when the number of selected sensors $k$ reaches a predefined number $p$. 
In contrast, the group greedy method reserves $L_{\rm{\max}}$ sets in each step and considers index candidates for $L_{\rm{\max}}$ sets obtained in the preceding step, while also evaluating a single objective function. %Here, $L_{\rm{\max}}$ is the number of reserved sets of sensors. 
The NMG method evaluates the index candidates for $L_{\rm{\max}}$ sets obtained in the preceding step, similar to the group greedy method. However, it reserves $L_{\rm{\max}}$ sets in each step based on the nondominated solutions of the MOP and evaluates the index candidates by considering the multi-objectives simultaneously. 

% Nondominated sortingの説明
In the proposed method, an efficient nondominated sort using the sequential search strategy (ENS-SS) \cite{zhang2015efficient} is used for obtaining the nondominated solutions. Note that the maximization problems in \eqref{eq:obj_det} and \eqref{eq:obj_eig} are transformed to minimization problems by inverting the sign of the objective functions when applying ENS-SS to them. 
The method first sorts all candidates in ascending order of the index of one of the objective functions. 
%(In the present study, the sorting uses the index of D-optimality.) 
The first set in the list of sorted candidates is assigned to the nondominated solutions (Pareto-optimal front) of rank $1$. The other candidates are then sequentially compared with the sets that have already been assigned to rank $1$ in terms of the other objective functions. 
%(In the present study, the indices of A- and E-optimality.) 
If there is no solution dominating the candidate, it is assigned to rank $1$. After the assignments of all candidates to the front of rank $1$ are checked, the succeeding rank is sequentially checked using the remaining candidates. 
See reference \cite{zhang2015efficient} for details of the algorithm of the ENS-SS. 

\begin{algorithm}[ht]
	\caption{Nondominated-solution-based Multi-objective Greedy Method} \label{alg:nmg}
	\begin{algorithmic}[1]
%    \STATE \textbf{Input:} the sensor-candidate matrix $\mathbf{U}\in\mathbb{R}^{n \times r}$, the number of sensors $p\in\mathbb{N}$, the number of sensor sets to be selected $L_{\rm{\max}}\in\mathbb{N}$
    \STATE \textbf{Input:} the number of the target set $p\in\mathbb{N}$, the number of sets to be selected in each step $L_{\rm{\max}}\in\mathbb{N}$, the set functions $f_i$, and the number of candidate indices $n$
    \STATE \textbf{Output:} a family of $p$ indices of $L_{\rm{max}}$ sets $\mathfrak{S}_{p} = \left\{\mathcal{S}^{1}_{p}, \dots, \mathcal{S}^{L_{\rm{max}}}_{p}\right\}$
    \STATE Set candidates of indices $\,\mathcal{S} := \left\{ 1, \hdots, n\right\}$;\\
    \STATE $\mathfrak{S}_{i}\leftarrow \emptyset \, (i \in \{1,\hdots,p\})$;\\
    % \STATE Set $k \leftarrow 1$;
    \FOR{$k=1, \dots, p$}%\FOR{ $k\leq p$ }
        \FOR{$l=1, \dots, L_{\rm{\max}}$}
            \FOR{$j = 1, \dots, v$}
                \STATE Calculate the values of objective functions for potential sets $f_j  \left(\mathcal{S}^{l}_{k-1}\cup \{ i \} \right)$ for $\forall\, {i\, \in\, \mathcal{S}\, \backslash\, \mathcal{S}^{l}_{k-1}}$;\\
            \ENDFOR
        \ENDFOR
        \STATE Eliminate duplication of potential sets;
        \STATE $rank \leftarrow 1$;
        \WHILE{true}
            \STATE $\mathfrak{F}_{k}[rank] \leftarrow$ the sets assigned to the $rank$--th front of the potential setes using ENS-SS;\\
            \STATE $L_{k}[rank] \leftarrow$ the number of sets in $\mathfrak{F}_{k}[rank]$;
            \IF{$\sum_{i=1}^{rank} L_{k}[i] \geq L_{\rm{max}}$}
                % \STATE $\mathfrak{S}_{k} \leftarrow \mathfrak{F}[i] \, (i \in \{1,\hdots,rank\})$
                \STATE $L_{k}' = L_{\rm{max}}-\sum_{i=1}^{rank-1}L_{k}[i]$;
                \STATE $\Delta \mathfrak{S}_{k} \leftarrow L_{k}'$ the sets selected using crowding distance and random selection from $\mathfrak{F}_{k}[rank]$;\\
                \STATE $\mathfrak{S}_{k} \leftarrow \mathfrak{S}_{k} \cup \Delta \mathfrak{S}_{k}$;\\
                % \STATE $\mathfrak{S}_{k} \leftarrow \mathfrak{S}_{k} \cup L'$ sensor sets selected using crowding distance and random selection from $\mathfrak{F}[rank]$;\\
                \STATE \textbf{break};
            % \ELSIF{$\sum_{i=1}^{rank}|L[i]|>L_{\rm{max}}$}
            %     % \STATE $\mathfrak{S}_{k} \leftarrow \mathfrak{F}[i] \, (i \in \{1,\hdots,rank-1\}) \, \cup$ sensor sets selected using crowding distance and random selection from $\mathfrak{F}[rank]$ s.t. $|\mathfrak{S}_{k}|=L_{\rm{max}}$;\\
            %     \STATE $\mathfrak{S}_{k} \leftarrow \mathfrak{S}_{k} \cup$ (sensor sets selected using crowding distance and random selection from $\mathfrak{F}[rank]$ s.t. $|\mathfrak{S}_{k}|=L_{\rm{max}}$);\\
            %     \STATE \textbf{break};
            \ELSE
                \STATE $\mathfrak{S}_{k} \leftarrow \mathfrak{S}_{k} \cup \mathfrak{F}_{k}[rank] $;
                \STATE $rank \leftarrow rank+1$;%$rank++$;
            \ENDIF
        \ENDWHILE
        %===
        \STATE $k \leftarrow k+1$;
    \ENDFOR
    \RETURN $\mathfrak{S}_{p}$;%\RETURN $\mathcal{S}_{1}, \dots, \mathcal{S}_{p}$;
	\end{algorithmic}
\end{algorithm}	

%解の個数の固定・crowding distanceに従うselection
In the present study, the number of reserved sets in each step is fixed to $L_{\rm{\max}}$. Thus, the nondominated sorting is terminated when the number of sets assigned to fronts reaches or exceeds $L_{\rm{\max}}$ and an explosive increase of candidates is prevented. The nondominated sorting can also be terminated when the rank of fronts to be checked reaches a predefined number, or it can continue until all candidates are assigned to fronts.
In addition, a candidate selection approach based on crowding distance is incorporated when selecting the required number of sets from the nondominated solutions.
The crowding distance, relative distance of each solution in a front, is often used in algorithms for solving MOPs.
The diversity of nondominated solutions was reported to be maintained by using this approach\cite{deb2002fast}. 
If the total number of sets assigned to fronts exceeds $L_{\rm{\max}}$ when the nondominated sorting is terminated, then the required number of sets needs to be selected from the last front.
The crowding distance of each set in the last front is computed, and then as many sets as required are selected in descending order of the crowd distance.
The crowding distance takes infinite values for sets that minimize or maximize any of objective functions in the last front.
The sets that do not minimize and maximize any objective functions are assigned to finite crowding distance values.
For details of the crowding distance, see \cite{deb2002fast}.
The order of the selection is not seriously considered among sets with infinite values of the crowding distance.
In the present study, we firstly select sets that minimize (maximize) any of objective functions for minimization (maximization) problems so that the NMG method becomes a relaxed greedy method that is defined later.
The relaxed greedy method has a preferable characteristic for optimization problems of monotone submodular or supermodular objective functions.
% In the present study, the sets of sensors with infinite crowding distance, which indicates that the value of either the D-, A-, and E-optimal indices of the sensor set is highest or lowest, are preferentially selected, and sensor sets randomly selected from the remaining candidates are added until the number of sensor sets reaches $L_{\rm{\max}}$.

Note that the ENS-SS algorithm has a best and worst case time complexity of $\mathcal{O}(vw\sqrt{w})$ and $\mathcal{O}(vw^2)$, respectively\cite{zhang2015efficient}. Here, $v$ and $w$ denote the number of objective functions and solutions to be sorted ($w=nL_{\rm{\max}}$), respectively. However, in the proposed method, the time complexity of the nondominated sorting using the ENS-SS is estimated to be $\mathcal{O}(vwL_{\rm{\max}})=\mathcal{O}(vnL^2_{\rm{\max}})$ by terminating the process based on the number of reserved sets, as explained below. This is typically less than the best time complexity of the ENS-SS algorithm since $n >> L_{\rm{\max}}$, e.g., $n=1000$ and $L_{\rm{\max}}=10$.

In this study, this NMG method including ENS-SS algorithm is applied to the sensor selection based on the multiple optimal experimental design. Here, the D-, A-, and E-optimality indices defined as \eqref{eq:obj_det}, \eqref{eq:obj_tr}, and \eqref{eq:obj_eig} are employed as the objective functions considered in the NMG method. \color{\ReviewerA}The purpose of using D-, A-, and E-optimization indices in the NMG method is that D-, A-, and E-optimization indices are objective functions suitable for improvement in the performance by the NMG method. With regard to D- and A-optimality, unified formulations for both under and oversampling situations of division-into-cases function in \eqref{eq:obj_det} and \eqref{eq:obj_tr} are proved to be submodular\cite{saito2021determinant,nakai2021effect}. The performance of the NMG method is guaranteed by a lower bound in a maximization problem of a monotone submodular objective function, as discussed from a theoretical perspective in Section~\ref{sec:theory}. In addition, the performance of the E-optimality-based objective function, the unified formulation of which is not submodular, is expected to improve due to the strong relationship between A- and E-optimality. It will also described in the last paragraph in Section~\ref{sec:theory}.\color{black}

\color{\ReviewerA}
\section{Theoretical Characteristics}
\label{sec:theory}
In this section, we define a class of the relaxed greedy methods that includes the pure greedy method, the group greedy method, and the proposed NMG method, and discuss the theoretical characteristics of the relaxed greedy methods. After that, expected characteristics of the NMG methods for the MOP for (\ref{eq:obj_det})--(\ref{eq:obj_eig}) are described.
We deal with maximization problems mainly in this section.
Theoretical results for the maximization problems can be naturally applied to equivalent minimization problems.

The pure greedy method and the group greedy method generate a sequence of sets of sets based on recursive maximization of a function to be maximized.
We define a class of methods that generate such a sequence as follows:
\begin{dfn}
For any $n \in \mathbb{N}$, some $p \in \mathbb{N}$ so that $p \le n$, and the finite set $\mathcal{S} \coloneqq \{ 1, 2, \dots, n \}$, a method $\mathcal{M}$ generates a sequence $(\mathfrak{S}_k \mid \forall \mathcal{S}_k \in \mathfrak{S}_k, \, \mathcal{S}_k \subset \mathcal{S}, \, \abs{\mathcal{S}_k} = k)_{k = 1, \dots, p}$.
For a set function $f: 2^\mathcal{S} \to \mathbb{R}$, we call $\mathcal{M}$ a relaxed greedy method (with respect to $f$) if there is $(\mathcal{S}_k^* \mid \mathcal{S}_k^* \in \mathfrak{S}_k)_{k = 1, \dots, p}$ that satisfies
% \begin{align} \label{eq: relaxed_greedy}
%     & %\left\{
%     \begin{array}{ll}
%         \mathcal{S}_k^* \in \argmax_{i \in \mathcal{S}} f(\{ i \}), & k = 1, \\
%         \mathcal{S}_k^* \in \argmax_{\mathcal{S}_{k - 1} \in \mathfrak{S}_ {k - 1}, \, i \in \mathcal{S} \backslash \mathcal{S}_{k-1}} f(\mathcal{S}_{k - 1} \cup \{ i \}), & k = 2, \dots, p.
%     \end{array} %\right.
% \end{align}
% \begin{align} \label{eq: relaxed_greedy}
%         \mathcal{S}_k^* \in \color{\ReviewerBB}\argmax_{\mathcal{S}_k' = \mathcal{S}_{k-1} \cup \{i\}, \mathcal{S}_{k - 1} \in \mathfrak{S}_ {k - 1}, \, i \in \mathcal{S} \backslash \mathcal{S}_{k-1}} \color{black}f(\mathcal{S}_k'), \quad k = 1, \dots, p,
% \end{align}
\color{\ReviewerBB}
\begin{align} \label{eq: relaxed_greedy}
    \mathcal{S}_k^* \in
    \argmax_{\mathcal{S}_k' \in \mathfrak{S}_k'} f(\mathcal{S}_k'),
    \quad k = 1, \dots, p,
\end{align}
where
\begin{align} \label{eq: set_of_potential_sets}
    \mathfrak{S}_k' =
    \{\mathcal{S}_{k - 1} \cup \{ i \} \mid \mathcal{S}_{k - 1} \in \mathfrak{S}_ {k - 1}, \, i \in \mathcal{S} \backslash \mathcal{S}_{k-1} \},
\end{align}
\color{black}
and $\mathfrak{S}_0 = \{\emptyset \}$.
\end{dfn}
The relaxed greedy method can be interpreted as generating $(\mathfrak{S}_k)_{k = 1, \dots, p}$ inductively.
The relaxed greedy method temporarily computes a set $\mathfrak{S}_k'$ of potential sets, and then, extracts $\mathfrak{S}_k \subset \mathfrak{S}_k'$ so that $\mathfrak{S}_k$ includes the optimal $\mathcal{S}_k^*$ for all the potential sets.
Note that the relaxed greedy method in the present study differs from the relaxed greedy algorithm in \cite{devore1996some}.

Some types of extended greedy methods are included in the class of relaxed greedy methods.
For example, the pure greedy method is apparently a relaxed greedy method because it generates a sequence $(\{ \mathcal{S}_k^* \})_{k = 1, \dots, p}$.
The group greedy method is also a relaxed greedy method because it generates a sequence $(\mathfrak{S}_k)_{k = 1, \dots, p}$, where $\mathfrak{S}_k$ is referred as $(\mathcal{S}_{k, l_i})_{l_i}$ in \cite{jiang2019group}, so that $\mathfrak{S}_k$ includes the optimal $\mathcal{S}_k^*$ that satisfies (\ref{eq: relaxed_greedy}).
These facts are summarized as the following theorem:
\begin{thm}
The pure greedy method and the group greedy method are in the class of relaxed greedy methods.
\end{thm}
%
%\begin{dfn}
%The single objective relaxed greedy method for a set function which adopts the multiple sets in ($k-1$)th step, investigates the $k$th potential sets based on  and selects the $k$th multiple sets including the best set in the range investigated is defined as the relaxed greedy methods.
%\end{dfn}
%\begin{thm}
%The group greedy methods is a class of relaxed greedy methods.
%\end{thm}
%\begin{thm}
%The nondominated-solution-based multi-objective greedy methods is a class of relaxed greedy methods.
%\end{thm}
%
In addition to the pure greedy and group greedy methods, the NMG method that we propose is also conditionally in the class of a relaxed greedy method:
\begin{thm}
The nondominated-solution-based multi-objective greedy method is a relaxed greedy method with respect to each of its objective functions if $L_{\max} \ge v$.
\end{thm}
\begin{proof}%\begin{IEEEproof}
We just need to show that, for any $j \in \{1, \dots, v\}$ and $k \in \{1, \dots, p\}$, $\mathfrak{S}_{k}$ contains at least one element of
\color{\ReviewerBB}
\begin{align}
    \mathfrak{S}_{j, \, k}^* \coloneqq
    \argmax_{\mathcal{S}_{k}' \in \mathfrak{S}_k'}
    f_j(\mathcal{S}_{k}'),
\end{align}
where $\mathfrak{S}_k'$ is the set of potential sets, which is defined in (\ref{eq: set_of_potential_sets}).
% \begin{align}
%     \mathcal{S}_{k}' \in
%     \left\{
%     \mathcal{S}_{k-1} \cup \{i\} \mid \mathcal{S}_{k - 1} \in \mathfrak{S}_ {k - 1}, \, i \in \mathcal{S} \backslash \mathcal{S}_{k-1}
%     \right\}.
% \end{align}
\color{black}
In the NMG method, $f_j  \left(\mathcal{S}_{k-1}\cup \{ i \} \right)$ is computed for all $\mathcal{S}_{k-1} \in \mathfrak{S}_{k-1}$ and $i \in \mathcal{S} \backslash \mathcal{S}_{k-1}$, and each potential set $\mathcal{S}_{k-1}\cup \{ i \}$ is assigned to one of nondominated fronts according to the value $f_j  \left(\mathcal{S}_{k-1}\cup \{ i \} \right)$ in the $k$th step.
All sets in $\mathfrak{S}_{j, \, k}^*$ are nondominated solutions, and therefore, they are assigned to the first front $\mathfrak{F}_{k}[1]$, that is, $\mathfrak{S}_{j, \, k}^* \subset \mathfrak{F}_{k}[1]$.
If the cardinal number of $\mathfrak{F}_{k}[1]$ is not greater than $L_{\max}$, i.e. $L_{k}[1] \le L_{\max}$, $\mathfrak{S}_{j, \, k}^* \subset \mathfrak{S}_k$ because $\mathfrak{F}_{k}[1] \subset \mathfrak{S}_k$.
If $L_{k}[1] > L_{\max}$, $\mathfrak{S}_k$ is obtained by selecting $L_{\max}$ sets from $\mathfrak{F}_{k}[1]$ according to the crowding comparison.
In our crowding comparison, first $v$ sets in $\mathfrak{S}_k$ are selected so that they maximize each of $v$ objective functions if $L_{\max} \ge v$.
Hence, at least one element of $\mathfrak{S}_{j, \, k}^*$ is contained in $\mathfrak{S}_k$ for any $j$ and $k$ if $L_{\max} \ge v$.
% First $L_{\max}$ sets sorted in descending order of their crowding distances are selected in the crowding distance comparison, and the crowding distance takes infinity only for sets that minimize or maximize any of $v$ objective functions in $\mathfrak{F}_{k-1}[1]$.
\end{proof}%\end{IEEEproof}
The relaxed greedy methods excluding the pure greedy method increase the number of sets for the maximization search, and the sets obtained by the relaxed greedy methods are usually better than or equal to those obtained by the pure greedy method as discussed in the present and previous papers. However, this dose not always stand. 
%\color{\ReviewerBB}It is noted that this dose not always stand. The pure greedy method gives better sensor sets than those obtained by the relaxed greedy methods in some cases. See the Appendix for a proof. \color{black}

\color{\ReviewerBB}
\begin{rem} \label{thm: relaxed}
The results of the relaxed greedy methods are not always better than nor equal to the pure greedy method.
\end{rem}
We show an example.
% \begin{thm} %\label{thm: relaxed}
% The results of the relaxed greedy methods are not always better than nor equal to the pure greedy method.
% \end{thm}
The 
%illustrative
example of the sensor problem with $r=2$, $n=5$, $p=3$, and  the group greedy method with $L_{\rm{max}}=2$ is adopted for this proof. The maximization of the D-optimality function in (\ref{eq:obj_det}) is considered where the sensor candidate matrix $\mathbf{U}$ is given as follows:
\begin{eqnarray}
\mathbf{U}=
\left[
\begin{array}{c}
\mathbf{u}_1 \\\mathbf{u}_2 \\\mathbf{u}_3 \\\mathbf{u}_4 \\\mathbf{u}_5 \\
\end{array}
\right]
=
\left[
\begin{array}{cc}
1.1000 & 0.0000 \\-0.5250 & 0.9093 \\ 0.7499 & 0.6616 \\ -0.7742 &-0.6329 \\ 0.6689 & 0.7434\\
\end{array}
\right].
\end{eqnarray}
%Those sensor vectors are also shown in Fig. . 
Here, $\|\mathbf{u}_1\|>\|\mathbf{u}_2\|>\|\mathbf{u}_3\|\sim\|\mathbf{u}_4\|\sim\|\mathbf{u}_5\|$. 
The determinant of the Fisher information matrix of the observation of two and three sensor combinations is summarized in Table \ref{table:det_FIM}. 
In this case, the sensor selection processes of the pure greedy method and the group greedy method which is one of the relaxed greedy methods are presented in Tables~\ref{table:proc_pureG} and 
\ref{table:proc_groupG}, respectively. This example shows that the pure greedy method gives the best sensor sets while the group greedy method does not. This is not often the case and the relaxed greedy methods usually shows better performance than the pure greedy method, but this example presents such an intuitive superiority is not always ensured.
\begin{center}
\begin{table}[!tbp]
% \begin{threeparttable}[ht]
    \small
    \centering
    \caption{\color{\ReviewerBB}The determinant of Fisher information matrix of two and three sensor combinations} 
    \label{table:det_FIM}
    \begin{tabular}{cc}
        \hline\hline
        sensor combination & determinant of FIM\\
        \hline
        \{1,2\} & 1.0005\\
        \{1,3\} & 0.5296\\
        \{1,4\} & 0.4847\\
        \{1,5\} & 0.6687\\
        \{2,3\} & 1.0593\\
        \{2,4\} & 1.0739\\
        \{2,5\} & 0.9970\\
        \{3,4\} & 0.0014\\
        \{3,5\} & 0.0132\\
        \{4,5\} & 0.0232\\        
        \hline
        \{1,2,3\} & 2.5894\\
        \{1,2,4\} & 2.5591\\
        \{1,2,5\} & 2.6662\\
        \{1,3,4\} & 1.0157\\
        \{1,3,5\} & 1.2115\\
        \{1,4,5\} & 1.1765\\
        \{2,3,4\} & 2.1346\\
        \{2,3,5\} & 2.0695\\
        \{2,4,5\} & 2.0941\\
        \{3,4,5\} & 0.0378\\        
        \hline\hline
    \end{tabular}
% \end{threeparttable}
\end{table}
\end{center}
\begin{center}
\begin{table}[!tbp]
% \begin{threeparttable}[ht]
    \small
    \centering
    \caption{\color{\ReviewerBB}Process of the pure greedy method.} 
    \label{table:proc_pureG}
    \begin{tabular}{cc}
        \hline\hline
        $k=1$ &  \\
        potential choice & \{1\}, \{2\}, \{3\}, \{4\}, \{5\} \\
        sensor selection of best configuration & $\mathcal{S}_1^*  = $\{1\}\\
        \hline
        $k=2$ &  \\
        potential choice & \{1,2\}, \{1,3\}, \{1,4\}, \{1,5\}\\
        sensor selection of best configuration & $\mathcal{S}_2^* = $\{1,2\}\\
        \hline        
        $k=3$ &  \\
        potential choice & \{1,2,3\}, \{1,2,4\}, \{1,2,5\}\\
        sensor selection of best configuration & $\mathcal{S}_3^* = $\{1,2,5\}\\
    \hline\hline
    \end{tabular}
% \end{threeparttable}
\end{table}
\end{center}
\begin{center}
\begin{table}[!tbp]
% \begin{threeparttable}[ht]
    \small
    \centering
    \caption{\color{\ReviewerBB}Process of the group greedy method ($L_{\max}$=2) which is one of the relaxed greedy methods. Here, $\mathcal{S}_{k,l}$ represents the $l$th best configuration in the $k$th step.} 
    \label{table:proc_groupG}
    \begin{tabular}{cc}
        \hline\hline
        $k=1$ &  \\
        potential choice & \{1\}, \{2\}, \{3\}, \{4\}, \{5\} \\
        best $L_{\max}$ configuration & $\mathfrak{S}_1 = \left\{ \{1\}, \, \{2\}\right\}$\\
        sensor selection & $\mathcal{S}_1^* = \{ 1 \}$\\
        \hline
        $k=2$ &  \\
        potential choice & \{1,2\}, \{1,3\}, \{1,4\}, \{1,5\}, \{2,3\}, \{2,4\}, \{2,5\}\\
        best $L_{\max}$ configuration & $\mathfrak{S}_2 = \left\{ \{2,4\}, \, \{2,3\} \right\}$ \\
        sensor selection & $\mathcal{S}_2^* = \{2,4\}$ \\
        \hline
        $k=3$ &  \\
        potential choice & \{1,2,3\}, \{1,2,4\}, \{2,3,4\}, \{2,3,5\},  \{2,4,5\}\\
        best $L_{\max}$ configuration & $\mathfrak{S}_3 = \left\{ \{1,2,3\}, \, \{1,2,4\} \right\}$\\
        sensor selection & $\mathcal{S}_3^* = \{1,2,3\}$\\ 
    \hline\hline
    \end{tabular}
% \end{threeparttable}
\end{table}
\end{center}
\color{black}
The pure greedy method is well known to give a desirable solution with the approximation rate $1 - 1/e$ if an objective function is monotone and submodular \cite{nemhauser1978analysis, krause2014submodular}.
Here, we define the monotonicity and the submodularity, respectively, as follows: 
\color{\ReviewerBB}
\begin{dfn}
A set function $f: 2^{\mathcal{S}} \to \mathbb{R}$ is monotone if, for any $\mathcal{S}_{\mathrm{a}}, \mathcal{S}_{\mathrm{b}} \subset \mathcal{S} $ with $\mathcal{S}_{\mathrm{a}} \subset \mathcal{S}_{\mathrm{b}}$, $f$ satisfies
\begin{align}
    f(\mathcal{S}_{\mathrm{a}}) \le f(\mathcal{S}_{\mathrm{b}}).
\end{align}
\end{dfn}
\begin{dfn}
A set function $f: 2^{\mathcal{S}} \to \mathbb{R}$ is called submodular if, for any  $\mathcal{S}_{\mathrm{a}}, \mathcal{S}_{\mathrm{b}} \subset \mathcal{S} $ with $\mathcal{S}_{\mathrm{a}} \subset \mathcal{S}_{\mathrm{b}}$ and $i \in \mathcal{S} \setminus \mathcal{S}_{\mathrm{b}}$, $f$ satisfies
\begin{align}
     f\bigl( \mathcal{S}_{\mathrm{a}} \cup \{ i \} \bigr) - f(\mathcal{S}_{\mathrm{a}})
 \ge f\bigl( \mathcal{S}_{\mathrm{b}} \cup \{ i \} \bigr) - f(\mathcal{S}_{\mathrm{b}}).
\end{align}
\end{dfn}
\color{black}
Although the previous Remark~\ref{thm: relaxed} implies that some of relaxed greedy methods sometimes give a worse solution than the pure greedy method, all relaxed greedy methods, as well as the pure greedy method, attain the approximation rate $1 - 1 / e$ for any monotone submodular objective function according to the following theorem:
% \begin{thm}
% The results f(Srgm) of the relaxed greedy methods for the maximization of the submodular function is greater than 0.63f(Sopt)
% \end{thm}
\begin{thm} \label{thm: rate}
Consider a monotone submodular function $f: 2^S \to \mathbb{R}$ such that $f(\emptyset) = 0$ and a sequence $(\mathcal{S}_k^* \mid \mathcal{S}_k^* \in \mathfrak{S}_k)_{k = 1, \dots, p}$ that is generated by a relaxed greedy method and satisfies (\ref{eq: relaxed_greedy}) for $f$.
Then, for a given integer $p$ and any $k \in \{ 1, \dots, p\}$,
\begin{align}
    f( \mathcal{S}_p^* ) \ge (1 - e^{- k / p})
    \max_{\mathcal{S}_p \subset \mathcal{S}, \, \abs{\mathcal{S}_p} \le p} f(\mathcal{S}_p).
\end{align}
In particular, $f( \mathcal{S}_p^* ) \ge (1 - 1 / e) \max_{\mathcal{S}_p \subset \mathcal{S}, \, \abs{\mathcal{S}_p} \le p} f(\mathcal{S}_p)$.
\end{thm}
\begin{proof}%\begin{IEEEproof}
    We only show that an important inequality holds for the proof of this theorem.
    Let us denote $\mathcal{S}_p^{\textrm{opt}} \subset \mathcal{S}$ that gives the maximum value $\max_{\mathcal{S}_p \subset \mathcal{S}, \, \abs{\mathcal{S}_p} \le p} f(\mathcal{S}_p)$.
    Because $f$ is monotone submodular, if we denote $\mathcal{S}_0^* = \emptyset$ for convenience, we have the following inequality for any $k \in \{ 1, \dots, p\}$:
    \begin{align}
        & f(\mathcal{S}_p^{\textrm{opt}}) - f(\mathcal{S}_{k-1}^*) \notag \\
        & \le f(\mathcal{S}_{k-1}^* \cup \mathcal{S}_p^{\textrm{opt}}) - f(\mathcal{S}_{k-1}^*) \notag \\
        & \le \sum_{i \in \mathcal{S}_p^{\textrm{opt}} \backslash \mathcal{S}_{k-1}^*}
        \Bigl(f(\mathcal{S}_{k-1}^* \cup \{ i \}) - f(\mathcal{S}_{k-1}^*)\Bigr) \notag \\
        & \le \abs{\mathcal{S}_p^{\textrm{opt}} \backslash \mathcal{S}_{k-1}^*} \left(
        \max_{i \in \mathcal{S}_p^{\textrm{opt}} \backslash \mathcal{S}_{k-1}^*} f(\mathcal{S}_{k-1}^* \cup \{ i \})
        - f(\mathcal{S}_{k-1}^*) \right).
        \label{eq: rate_proof1}
    \end{align}
    Because $( \mathcal{S}_k^* \in \mathfrak{S}_k )_{k = 1, \dots, p}$ satisfies (\ref{eq: relaxed_greedy}), the following inequality holds:
    \begin{align}
        & \max_{i \in \mathcal{S}_p^{\textrm{opt}} \backslash \mathcal{S}_{k-1}^*} f(\mathcal{S}_{k-1}^* \cup \{ i \}) \notag \\
        & \le \max_{i \in \mathcal{S} \backslash \mathcal{S}_{k-1}^*} f(\mathcal{S}_{k-1}^* \cup \{ i \}) \notag \\
        & \le \max_{\mathcal{S}_{k - 1} \in \mathfrak{S}_ {k - 1}, \, i \in \mathcal{S} \backslash \mathcal{S}_{k-1}} f(\mathcal{S}_{k - 1} \cup \{ i \})
        = f(\mathcal{S}_{k}^*), \quad \forall k.
        \label{eq: rate_proof2}
    \end{align}
    From (\ref{eq: rate_proof1}), (\ref{eq: rate_proof2}), and $\abs{\mathcal{S}_p^{\textrm{opt}} \backslash \mathcal{S}_{k-1}^*} \le p$, we have
    \begin{align}
        f(\mathcal{S}_p^{\textrm{opt}}) - f(\mathcal{S}_{k-1}^*)
        \le p \left(
        f(\mathcal{S}_{k}^*)
        - f(\mathcal{S}_{k-1}^*) \right), \quad \forall k.
        \label{eq: rate_proof3}
    \end{align}
    By using (\ref{eq: rate_proof3}), the remain of the proof can be completed in the same way as the proof for the pure greedy method.
    See, e.g., \cite{krause2014submodular} for the proof for the pure greedy method.
\end{proof}%\end{IEEEproof}

Although different relaxed greedy methods give different approximate solutions in general, Theorem~\ref{thm: rate} guaranties the same lower bounds for the approximated solutions in a maximization problem of a monotone submodular objective function.

Theorem~\ref{thm: rate}, unfortunately, does not guarantee quality of approximated solutions that are obtained by the NMG method for the optimization problem of D-, A-, and E-optimality functions in (\ref{eq:obj_det})--(\ref{eq:obj_eig}) because these division-into-cases functions are not proven to be submodular or supermodular. However, D- and A-optimality functions can be slightly modified to be submodular functions by adding $\epsilon \mathbf{I}$ to the matrix, i.e. $\mathbf{C}^{\mathsf{T}}\mathbf{C}+\epsilon \mathbf{I}$, in the objective functions in the oversampling situation and by using obtained unified objective function for both under and oversampling situations, as shown in \cite{saito2021determinant,nakai2021effect}, whereas $\epsilon$ is a hyperparameter. 
For a maximization problem of these modified functions, Theorem~\ref{thm: rate} can be applied. It should be noted that the objective function of D- and A-optimality functions in (\ref{eq:obj_det}) and (\ref{eq:obj_tr}) are limit of the unified submodular functions when the hyperparameter $\epsilon \rightarrow 0$. 

As we discussed in the previous paper, the objective function based on A-optimality has strong relationship with that based on E-optimality\cite{nakai2021effect}. The objective function of the A- and E-optimality-based greedy methods can be written for the oversampling situation as follows: 
\begin{align}
f_{\textrm{A}}&=\frac{1}{\lambda_1} + \frac{1}{\lambda_2} + \cdots + \frac{1}{\lambda_r}, \\
f_{\textrm{E}}&=\lambda_r,
\end{align}
where $\lambda_i$ is the $i$--th largest eigenvalue of $\mathbf{C}^{\top}\mathbf{C}$ and $\lambda_r=\lambda_{\text{min}}$. The increase in the minimum eigenvalue $\lambda_r$ is important for the A-optimality-based objective function because $1 / \lambda_r = 1/ f_{\textrm{E}}$ is the largest term in $f_{\textrm{A}}$. Hence, the increase in the E-optimality-based objective function is strongly demanded in the minimization of the A-optimality-based objective function. This fact was confirmed in the previous paper; the numerical experiments have shown that the A-optimality-based greedy method performs well in terms of maximizing $\lambda_r$. 
On the other hand, the E-optimality-based greedy method was shown not to effectively work for the maximization of $\lambda_r$ compared with the A-optimality-based greedy method, especially in the oversampling situation. This seems because the A-optimality-based objective function in (\ref{eq:obj_tr}) is a limit of the unified submodular function as discussed above whereas that of the E-optimality-based objective function in (\ref{eq:obj_eig}) is not submodular~\cite{nakai2021effect} even in the relaxed situation by adding $\epsilon \mathbf{I}$ to the matrix of the objective function. The connection to the submodularity above in the objective function possibly leads to the superior performance of the A-optimality-based relaxed greedy method. 
%あまりに性能がでないと書きすぎているのでコメント
%, though the A-optimality-based objective function in (\ref{eq:obj_tr}) is not proven to be submodular or supermodular nor the quality of solutions obtained by the greedy method using the A-optimality-based objective function in (\ref{eq:obj_tr}) is not strictly guaranteed. 
%上で書いたのでコメント
%Again, the unified formulations are slightly modified functions of (\ref{eq:obj_tr}) and (\ref{eq:obj_eig}) derived in both cases of undersampling and oversampling situations. The performance of the greedy method using the unified formulation of A-optimality-based objective function is guaranteed by the lower bound because of its submodularity. Note that the A-optimality-based objective function in (\ref{eq:obj_tr}) is not proven to be submodular or supermodular, and therefore, the quality of solutions obtained by the greedy method using the A-optimality-based objective function in (\ref{eq:obj_tr}) is not strictly guaranteed. 
Therefore, in the proposed NMG method that simultaneously evaluates A- and E-optimality using (\ref{eq:obj_tr}) and (\ref{eq:obj_eig}), the index of A-optimality improves due to the connection to the submodularity and Theorem~\ref{thm: rate}, and consequently, the index of E-optimality is expected to be improved significantly due to the strong relationship between the objective function based on A- and E-optimality. 
%上で書いたので不要
%Finally, the unified formulation of the D-optimality-based objective function in (\ref{eq:obj_det}) is proven to be submodular\cite{saito2021determinant}. 
%Again, Theorem~\ref{thm: rate} does not strictly guarantee quality of approximated solutions that are obtained by the NMG method for the optimization problem of D- and A-optimality functions in (\ref{eq:obj_det}) and (\ref{eq:obj_tr}), but for a maximization problem of these modified functions, Theorem~\ref{thm: rate} can be applied. 

\color{\ReviewerAA}It should be noted that the pseudo-greedy method using an alternative criterion for the improvement of E-optimality, which is computationally more efficient and effective than the conventional E-optimality-based greedy method defined in \eqref{eq:obj_eig}, has been proposed\cite{jiang2016sensor}. Here, the alternative criterion is to maximize the projection of the observation vector of $k$--th sensor onto the minimum eigenspace of ($k$-1)--th observation matrix $\mathbf{C}$. Although the combination of the NMG method and this objective function seems to further accelerate sensor optimization in terms of E-optimality, this point is out of the scope of the present study and left for the future study. \color{black}

\section{Results and Discussion}
\label{sec:res}
In this section, the NMG method is applied to the sensor selection problem for randomly generated systems. The performance of the method is evaluated in terms of the value of the objective functions using indices of D-, A-, and E-optimality.
The random sensor-candidate matrices, $\mathbf{U}\in \mathbb{R}^{n{\times}r}$, are set so that the component of the matrices is given by the Gaussian distribution of $\mathcal{N}(0,1)$, and the $p$ sensors are selected by the NMG and several previously proposed methods. The sensor selection methods investigated are listed in Table~\ref{table:method_list}. \reviewerB{The number of sensor candidates was set to $n=100$, 1000, and 10000}, and the number of latent state variables was set to $r=10$, and the number of reserved sets of sensors in each step is set to $L_{\rm{max}} = 5$, 10, 20, and 50.
We first compare the NMG method to three pure greedy methods that utilize a single objective function based on each index of D-, A-, and E-optimality. Next, the NMG method is compared to three group greedy methods that consider a single objective function based on each index of D-, A-, and E-optimality and that reserve a group of suboptimal sensor sets in each step. Finally, the relationship of the D-, A-, and E-optimality criteria is discussed based on the Pareto front obtained by the NMG method. 
The numerical experiments were conducted under the computational environment described in Table~\ref{table:comp_env}. %The MATLAB code is available on Github%\cite{nakai2021github}.
%=======================
\begin{center}
\begin{table}[!tbp]
% \begin{threeparttable}[ht]
    \small
    \centering
    \caption{Sensor selection methods investigated in this study} 
    \label{table:method_list}
    \begin{tabular}[t]{lll}
        \hline\hline
        Name & Objective function & Scheme\\%Algorithm
        \hline
        NMG & D, A and E & Nondominated-solution-based\\
        & & multi-objective greedy method\\
        \hline
        DG\cite{saito2021determinant} & D & ${p \le  r}$ : pure greedy method based on QR \\
        & & ${p>r}$ : pure greedy method\\
        \hline
        AG\cite{nakai2021effect} & A & pure greedy method\\
        \hline
        EG\cite{nakai2021effect} & E & pure greedy method\\
        \hline
        DGG & D & group greedy method\cite{jiang2019group}\\
        \hline
        AGG & A & group greedy method\cite{jiang2019group}\\
        \hline
        EGG & E & group greedy method\cite{jiang2019group}\\
        \hline\hline
    \end{tabular}
% \end{threeparttable}
\end{table}
\end{center}

%=======================
\begin{center}
\begin{table}[!tbp]
% \begin{threeparttable}[ht]
    \small
    \centering
    \caption{Computational environment} 
    \label{table:comp_env}
    \begin{tabular}[t]{ll}
        \hline\hline
        Processor information & Intel(R) Xeon(R) W-2295 CPU @ 3.00 GHz \\
        \hline
        Random access memory & 256 GB\\
        \hline
        Operating system & Ubuntu 20.04.2 LTS\\
        \hline
        System type & 64-bit operating system\\
        \hline
        Source code & MATLAB R2022a\\
        \hline\hline
    \end{tabular}
% \end{threeparttable}
\end{table}
\end{center}

\subsection{Comparison with Pure Greedy Method}
The three pure greedy methods used for comparison with the NMG method were the D-optimality-based greedy (DG) method, the A-optimality-based greedy (AG) method, and the E-optimality-based greedy (EG) method, as listed in Table~\ref{table:method_list}.

\begin{figure*}[htbp]
    \centering
    \includegraphics[width=17cm]{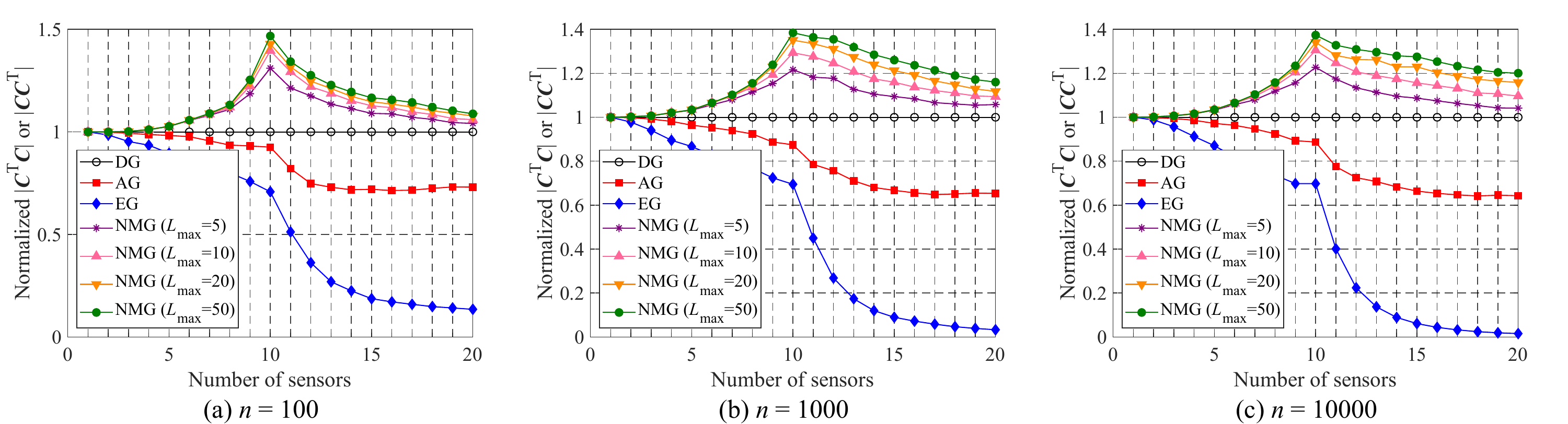}
    \caption {Normalized index of D-optimality defined as \eqref{eq:obj_det} against the number of sensors obtained by the NMG and pure greedy methods for random systems. The results are normalized by the value from the DG method.}
    \label{fig:NMG-PG_determinant}
\end{figure*}

\begin{figure*}[htbp]
    \centering
    \includegraphics[width=17cm]{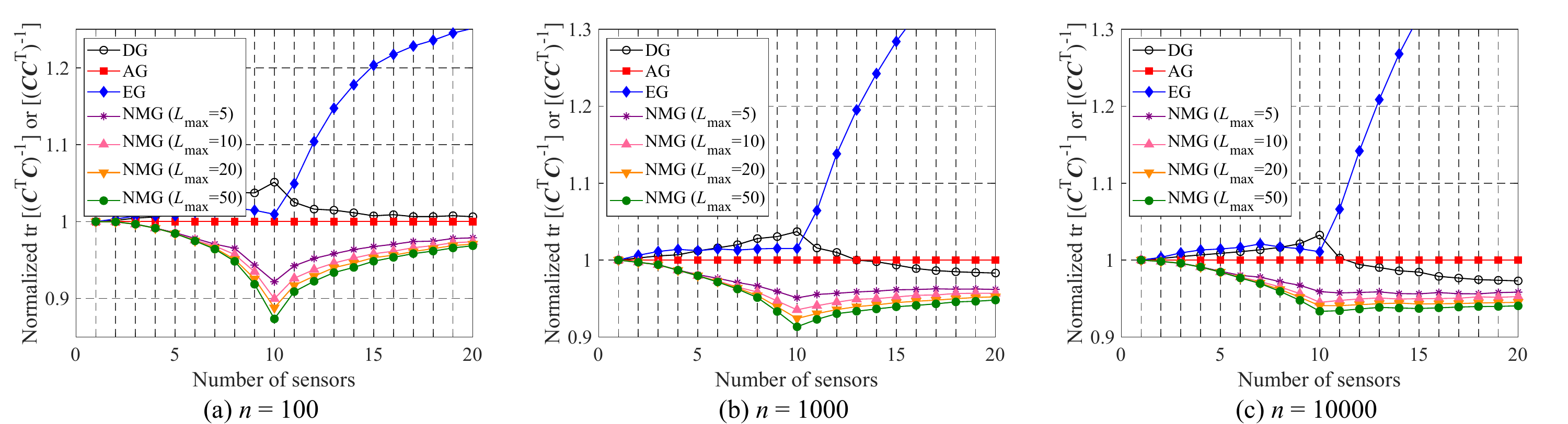}
    \caption{Normalized index of A-optimality defined as \eqref{eq:obj_tr} against the number of sensors obtained by the NMG and pure greedy methods for random systems. The results are normalized by the value from the AG method.}
    \label{fig:NMG-PG_trace}
\end{figure*}

\begin{figure*}[htbp]
    \centering
    \includegraphics[width=17cm]{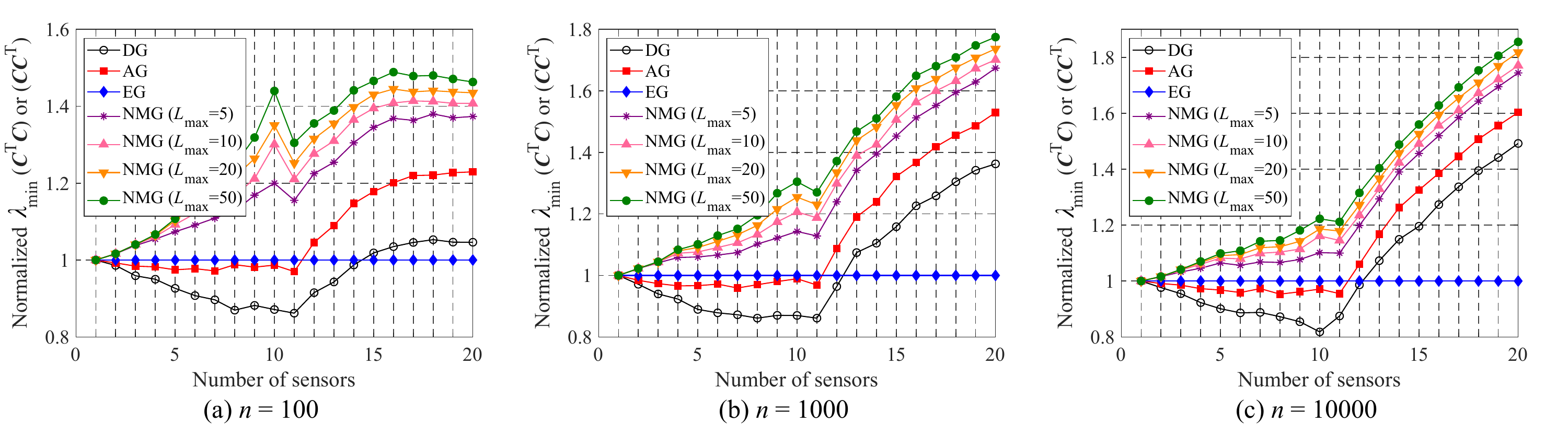}
    \caption{Normalized index of E-optimality defined as \eqref{eq:obj_eig} against the number of sensors obtained by the NMG and pure greedy methods for random systems. The results are normalized by the value from the EG method.}
    \label{fig:NMG-PG_eigenvalue}
\end{figure*}

\begin{figure}[htbp]
    \centering
    \includegraphics[width=3in]{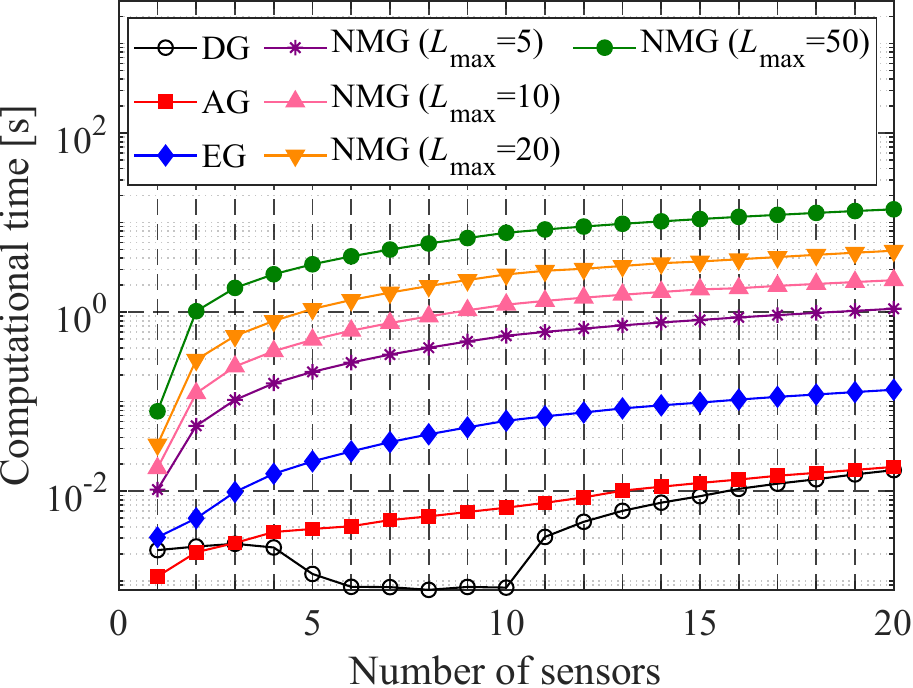}
    \caption{Sensor selection time against the number of sensors in the NMG method and pure greedy methods for random systems in the case of $n=1000$.}
    \label{fig:NMG-PG_time}
\end{figure}

Figs.~\ref{fig:NMG-PG_determinant}, \ref{fig:NMG-PG_trace}, and \ref{fig:NMG-PG_eigenvalue} show the relationship between the number of sensors and the three optimality indices for sensors selected by each method defined as \eqref{eq:obj_det}, \eqref{eq:obj_tr}, and \eqref{eq:obj_eig}. The values plotted in Figs.~\ref{fig:NMG-PG_determinant}, \ref{fig:NMG-PG_trace}, and \ref{fig:NMG-PG_eigenvalue} are normalized by the values obtained by the DG, AG, and EG methods. Note that the indices obtained by the NMG method correspond to the best values of reserved sets of sensors. Fig.~\ref{fig:NMG-PG_time} shows the computational time required for sensor selection. The results of the NMG, DG, AG, and EG methods are plotted together for comparison. The values plotted in Figs.~\ref{fig:NMG-PG_determinant}, \ref{fig:NMG-PG_trace}, \ref{fig:NMG-PG_eigenvalue}, and~\ref{fig:NMG-PG_time} are averages over numerical tests using 100 random sensor-candidate matrices.

Figs.~\ref{fig:NMG-PG_determinant}, \ref{fig:NMG-PG_trace}, and \ref{fig:NMG-PG_eigenvalue} reveal that the NMG method outperforms the pure greedy methods for all optimality indices in all the cases of $L_{max} = 5$, 10, 20, and 50 \reviewerB{and $n=100$, 1000, and 10000}. Fig.~\ref{fig:NMG-PG_determinant} shows that the index of D-optimality obtained by the NMG method is higher for any number of sensors relative to that obtained by the DG method, which has the highest value among the pure greedy methods. (See reference \cite{nakai2021effect} for details of the performance characteristics of the three pure greedy methods.) Figs.~\ref{fig:NMG-PG_trace} and \ref{fig:NMG-PG_eigenvalue} show that the indices of A- and E-optimality obtained by the NMG method is the lowest and highest, respectively, for any number of sensors relative to the values obtained by the pure greedy methods. This indicates that the NMG method is able to find sets of sensors superior to the pure greedy methods based on a single objective function in terms of D-, A-, and E-optimality. Furthermore, the indices of D-, A-, and E-optimality improve with increasing $L_{\rm{\max}}$. The high performance of the NMG method is mainly due to the increase in the search space that comes by reserving $L_{\rm{max}}$ solutions. It is also due to the simultaneous consideration of the various indices of optimality in the NMG method, which improves the performance of the greedy method for the index that is hardly maximized by the pure greedy method, i.e., the EG does not perform better than the DG and AG methods even in terms of E-optimality\cite{nakai2021effect}.

Fig.~\ref{fig:NMG-PG_time} shows that the NMG method requires more time than the pure greedy methods and that the computational time of the NMG method increases as $L_{\rm{\max}}$ increases. The difference between the NMG and pure greedy methods is mainly because of the nondominated sorting and evaluation of optimality indices, which require most of the computational time in the NMG method. The nondominated sorting utilizing crowding distance is conducted in each step of the sensor selection by the NMG method. In addition, the indices of three optimality of sensor candidates for $L_{\rm{\max}}$ sets of sensors are evaluated in the NMG method, while the pure greedy methods consider an index of a single optimality for one set of sensors. Furthermore, the computational time for the NMG method gradually increases with increasing $p$ because it chooses sensors in a step-by-step manner, as that for the pure greedy methods.

\subsection{Comparison with Group Greedy Method}
\begin{figure*}[htbp]
    \centering
    \includegraphics[width=17cm]{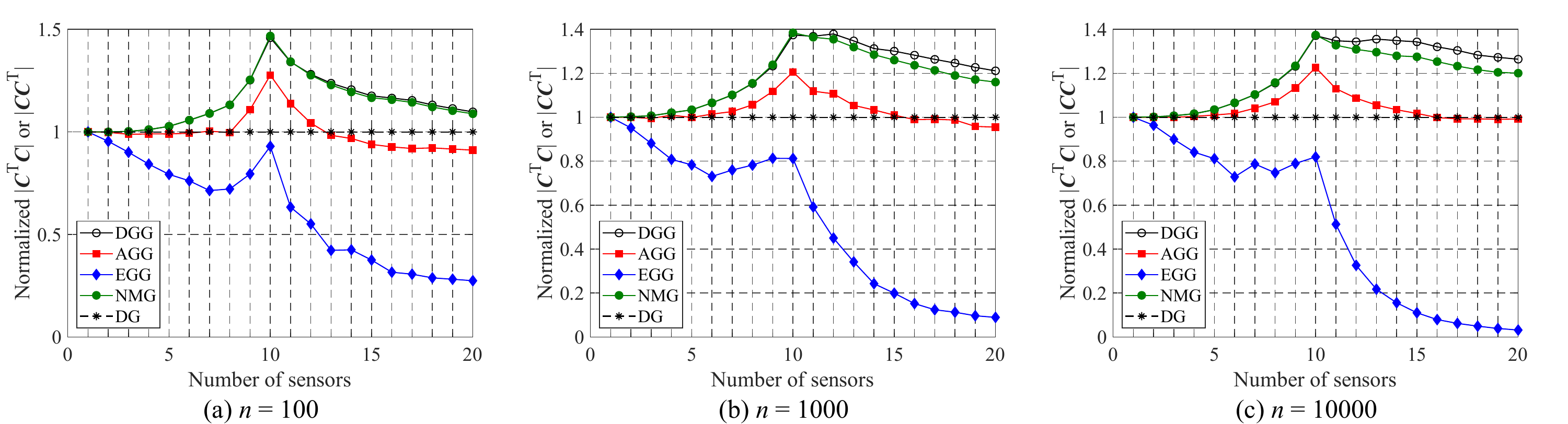}
    \caption {Normalized index of D-optimality defined as \eqref{eq:obj_det} against the number of sensors obtained by the NMG and group greedy methods for random systems in the case of $L_{\rm{\max}}=50$. The results are normalized by the value from the DG method.}
    \label{fig:NMG-GG_determinant}
\end{figure*}

\begin{figure*}[htbp]
    \centering
    \includegraphics[width=17cm]{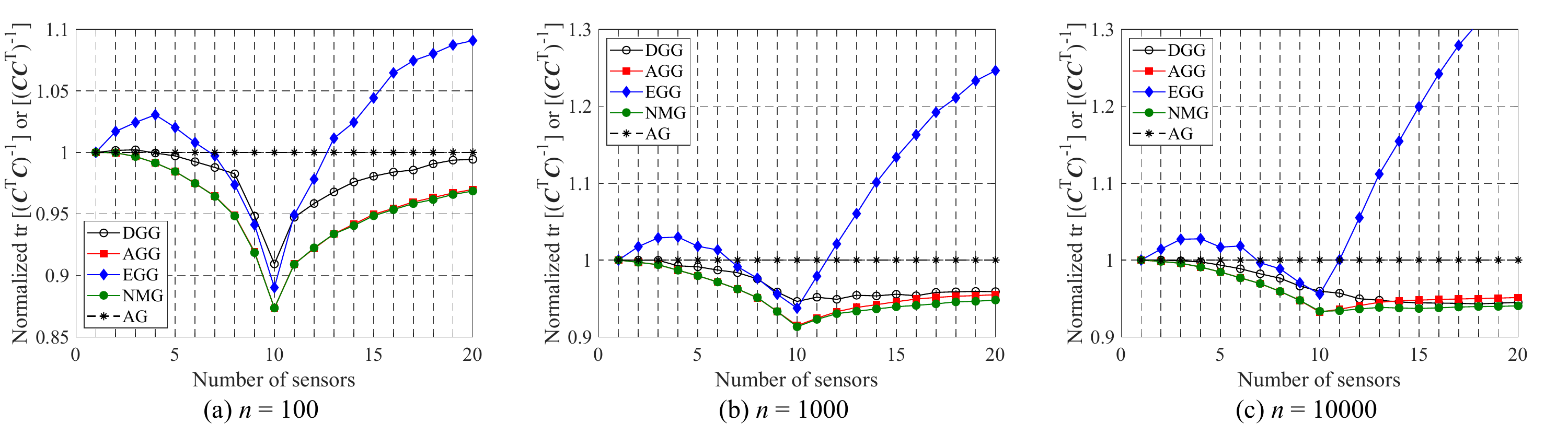}
    \caption{Normalized index of A-optimality defined as \eqref{eq:obj_tr} against the number of sensors obtained by the NMG and group greedy methods for random systems in the case of $L_{\rm{\max}}=50$. The results are normalized by the value from the AG method.}
    \label{fig:NMG-GG_trace}
\end{figure*}

\begin{figure*}[htbp]
    \centering
    \includegraphics[width=17cm]{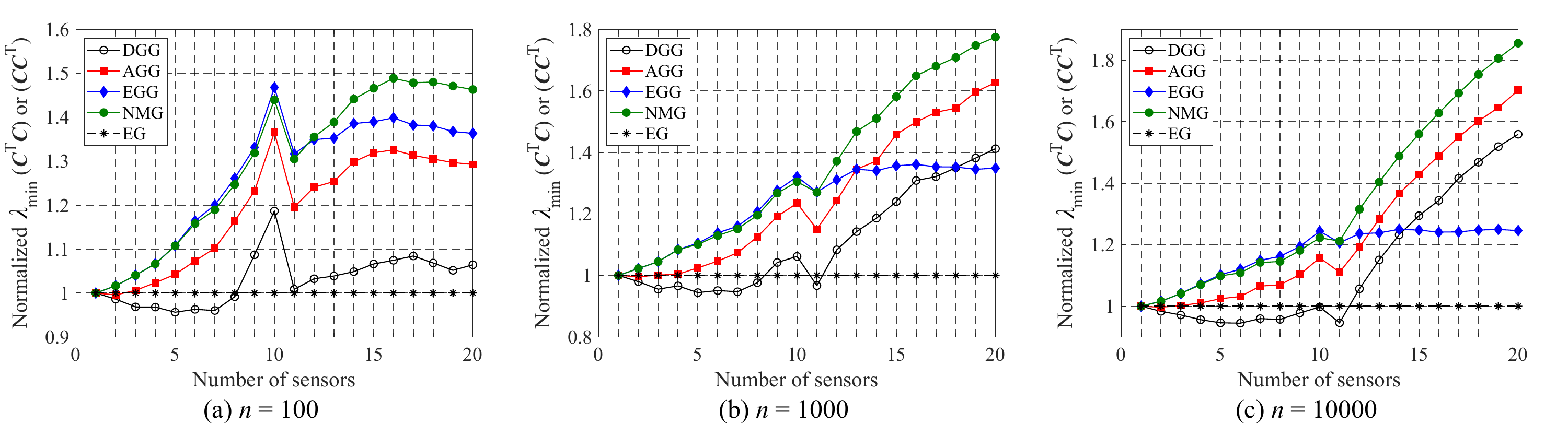}
    \caption{Normalized index of E-optimality defined as \eqref{eq:obj_eig} against the number of sensors obtained by the NMG and group greedy methods for random systems in the case of $L_{\rm{\max}}=50$. The results are normalized by the value from the EG method.}
    \label{fig:NMG-GG_eigenvalue}
\end{figure*}

\begin{figure}[htbp]
    \centering
    \includegraphics[width=3in]{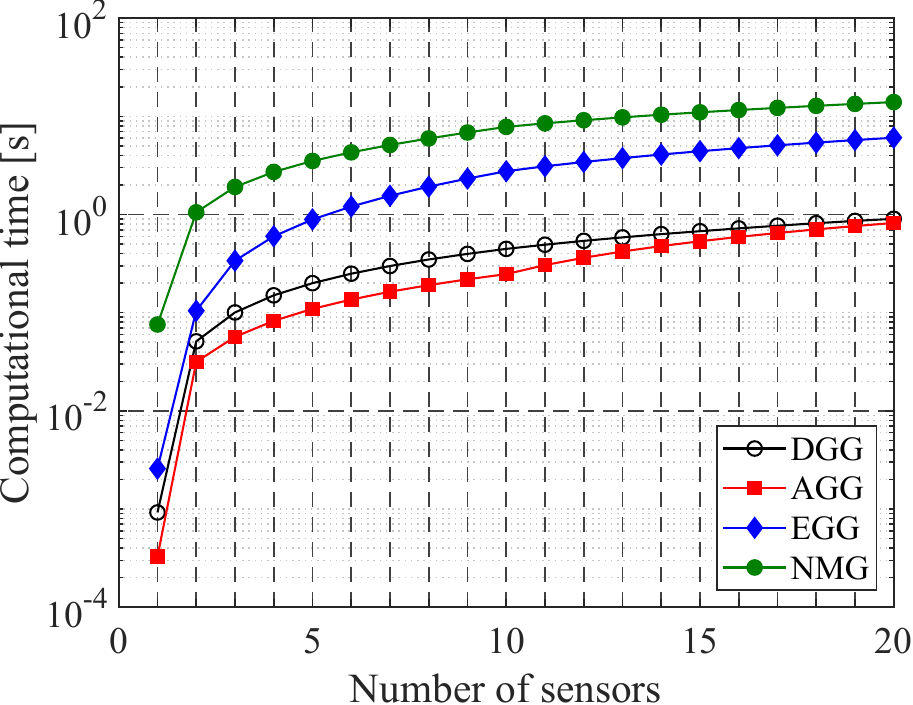}
    \caption{Sensor selection time against the number of sensors in the NMG method and group greedy methods for random systems in the case of $n=1000$ and $L_{\rm{\max}}=50$.}
    \label{fig:NMG-GG_time}
\end{figure}

The three group greedy methods used in the NMG comparison were the D-optimality-based group greedy (DGG) method, the A-optimality-based group greedy (AGG) method, and the E-optimality-based group greedy (EGG) method, as listed in Table~\ref{table:method_list}.
Figs.~\ref{fig:NMG-GG_determinant}, \ref{fig:NMG-GG_trace}, and \ref{fig:NMG-GG_eigenvalue} show the relationship between the number of sensors and the three optimality indices. The values plotted in Figs.~\ref{fig:NMG-GG_determinant}, \ref{fig:NMG-GG_trace}, and \ref{fig:NMG-GG_eigenvalue} are normalized by the values obtained by the DG, AG, and EG methods. Fig.~\ref{fig:NMG-PG_time} shows the computational time required for sensor selection. The values plotted in Figs.~\ref{fig:NMG-GG_determinant}, \ref{fig:NMG-GG_trace}, \ref{fig:NMG-GG_eigenvalue}, and~\ref{fig:NMG-GG_time} are averages over numerical tests using 100 random sensor-candidate matrices. The number of reserved sets of sensors $L_{\rm{\max}}$ for the NMG and GG methods was set to 50. Qualitative characteristics are confirmed to be the same in all the cases of $L_{\rm{\max}}=5$, 10, 20, and 50 in the present study. The number $L_{\rm{\max}}$ of reserved sets of sensors in the NMG and GG methods is set to the same as each other, and fair comparison is made with regard to the size of the search space, though the number of objective function evaluations by the NMG method is three times as large as that by the GG methods, 

Figs.~\ref{fig:NMG-GG_determinant}, \ref{fig:NMG-GG_trace}, and \ref{fig:NMG-GG_eigenvalue} reveal that \color{\ReviewerB}in all the cases of $n=100$, 1000, and 10000, \color{black} the NMG method outperforms the group greedy methods for the three optimality indices in nearly all cases of $p$. 
Figs.~\ref{fig:NMG-GG_trace} and \ref{fig:NMG-GG_eigenvalue} show the indices of A- and E-optimality obtained by the NMG method are the same as or lower than and  the same as or higher than those obtained by the group greedy methods for any number of sensors, respectively. In addition, Fig.~\ref{fig:NMG-GG_determinant} shows that the index of D-optimality obtained by the NMG method is as high as that obtained by the DGG method, which is the highest among the group greedy methods, in the case of $p \leq r$. These results indicate that the NMG method can ensure a diversity of solutions by increasing the search space, similar to the group greedy strategy. Furthermore, in most cases, the NMG method is able to find sets of sensors that are superior to those found by the group greedy methods based on a single objective function in terms of A- and E-optimality. This is because the NMG method reserves the nondominated solutions of D-, A-, and E-optimality.

In particular, \color{\ReviewerB}in all the cases of $n=100$, 1000, and 10000, \color{black} the index of E-optimality obtained by the NMG method is significantly superior to that obtained by the EGG method in the case of $p > r$, while it is nearly the same in the case of $p\leq r$, as shown in Fig.~\ref{fig:NMG-GG_eigenvalue}. The increasing trend of the index of E-optimality obtained by the NMG method is similar to that obtained by the AGG method when $p > r$. Moreover, the AG method is the most suitable for the maximization of the index of E-optimality among the pure greedy methods when $p$ exceeds $r$, as shown in Fig.~\ref{fig:NMG-PG_eigenvalue}. This has been reported to be due to the lack of submodularity in the E-optimality-based objective function\cite{nakai2021effect}. 
Thus, the high performance of the NMG method in terms of E-optimality is mainly due to considering the E-optimality-based objective function until $p \simeq r$, and considering the A-optimality-based objective function when $p > r$. \color{\ReviewerA}The results agree with the theoretical consideration of the performance improvement due to the relationship between the objective functions based on A- and E-optimality in Section~\ref{sec:theory}. \color{black}

Similarly, the high performance of the NMG method in terms of A-optimality is mainly due to considering the A-optimality-based objective function until $p \simeq r$, and considering the D-optimality-based objective function when $p$ exceeds $r$.
Fig.~\ref{fig:NMG-GG_trace} reveals that the index of A-optimality obtained by the NMG method is the same as that obtained by the AGG method until $p \simeq r$. On the other hand, as $p$ increases, it is superior to that obtained by the AGG method. %, while the trend of the change in the A-optimality index obtained by the NMG method is similar to that obtained by the DGG method. In addition, 
\color{\ReviewerB}Fig.~\ref{fig:NMG-PG_trace} indicates that the DG method tends to be superior to the AG method in terms of A-optimality in the oversampling situation as $n$ increases; the index of A-optimality obtained by the DG method is lower than that obtained by the AG method when $p$ exceeds 13 and 11 in the case of $n = 1000$ and 10000, respectively. Hence, evaluation of D-optimality possibly contributes to the performance of A-optimality obtained by the NMG method in the oversampling situation. \color{black}

At the same time, the index of D-optimality is lower for the NMG method than for the DGG method in the case of $p > r$. This is due to the fact that the DG method is the most suitable for the maximization of the index of D-optimality among the pure greedy methods for any number of sensors, as shown in Fig.~\ref{fig:NMG-PG_determinant}. The NMG method reserves $L_{\rm{\max}}$ nondominated solutions of D-, A-, and E-optimality, which contain solutions with worse performance for D-optimality and better performance for A- or E-optimality. On the other hand, the DGG reserves $L_{\rm{\max}}$ solutions considering only superiority with respect to D-optimality. Consequently, the sensor sets obtained by the NMG method are not as diverse as those obtained by the DGG method in terms of D-optimality in the range we investigated. It should be noted, however, that the NMG method can be expected to find better sets of sensors for D-optimality by increasing $L_{\rm{\max}}$, so that the diversity of sensor sets in terms of D-optimality is sufficiently ensured. 

Fig.~\ref{fig:NMG-GG_time} shows that the NMG method requires longer computational time than the group greedy methods. This is mainly due to the nondominated sorting utilizing crowding distance conducted in the NMG method, as explained in Section~\ref{sec:probandalg}. Furthermore, all indices of the three optimality are evaluated by the NMG method, while only an index of a single optimality is evaluated by the group greedy methods. 
Thus, the NMG method is able to find better sensor sets than the pure and group greedy methods, but with longer computational time. Note that the time complexity can be reduced to $\mathcal{O}\left(vnL^2_{\rm{\max}}\right)$ in the proposed method compared to the best time complexity of $\mathcal{O}\left(vnL_{\rm{\max}}\sqrt{nL_{\rm{\max}}}\right)$ of the ENS-SS algorithm, as explained in Section~\ref{sec:probandalg}. This seems to be a reasonable additional cost even for large-scale problems. 

\subsection{Pareto Font of Nondominated-solution-based Multi-objective Greedy Method}
In this section, the relationship among the D-, A-, and E-optimality criteria is discussed based on the Pareto front obtained by the NMG method. 

\begin{figure}[htbp]
    \centering
    \includegraphics[width=3in]{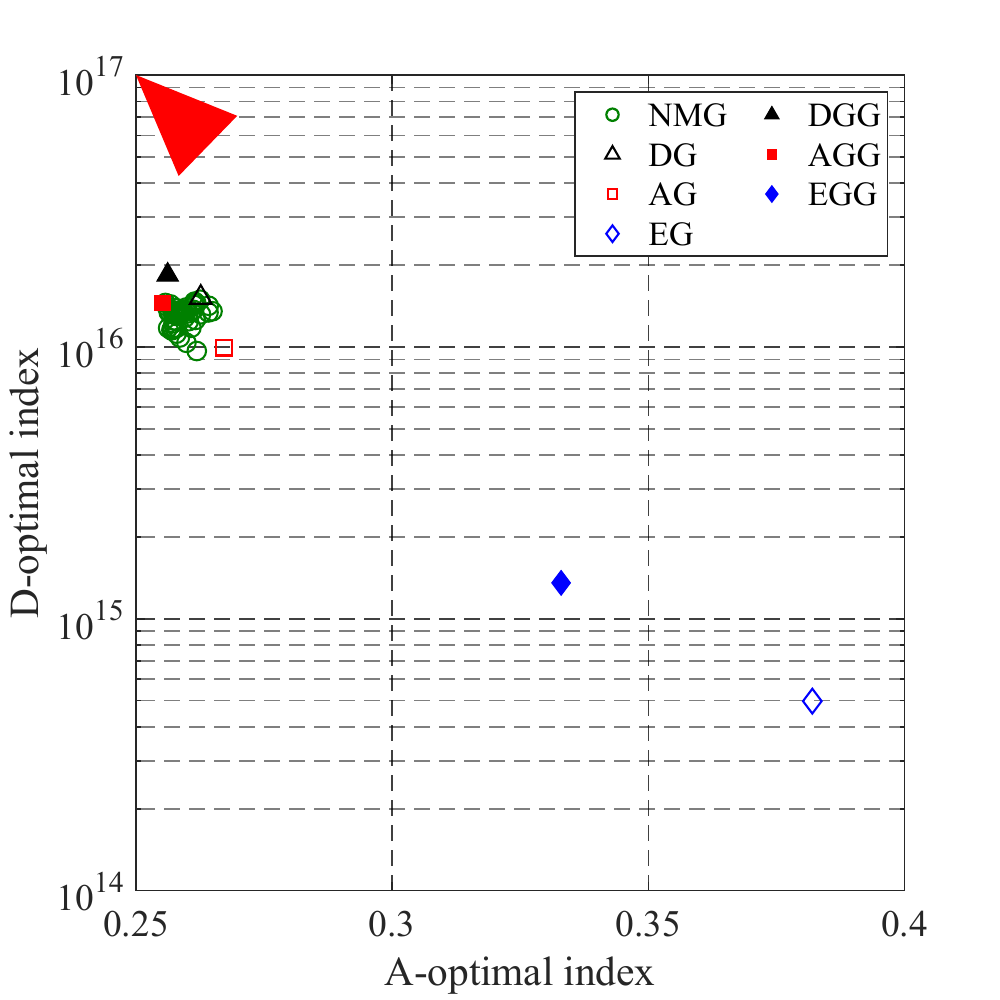}
    \caption {Relationship between the indices of D- and A-optimality for $p=20$, $n=1000$ and $L_{\rm{\max}}=50$.}
    \label{fig:ParetoDA}
\end{figure}

\begin{figure}[htbp]
    \centering
    \includegraphics[width=3in]{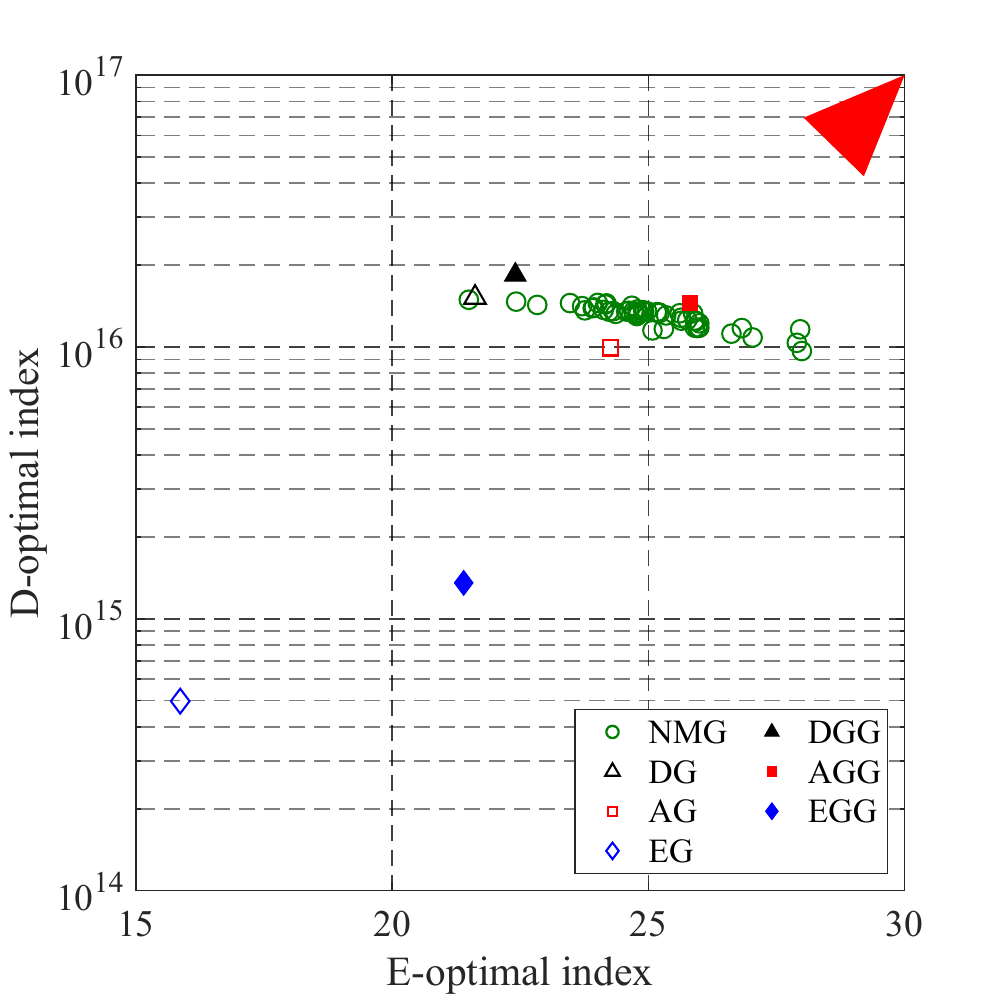}
    \caption {Relationship between the indices of on D- and E-optimality for $p=20$, $n=1000$ and $L_{\rm{\max}}=50$.}
    \label{fig:ParetoDE}
\end{figure}

\begin{figure}[htbp]
    \centering
    \includegraphics[width=3in]{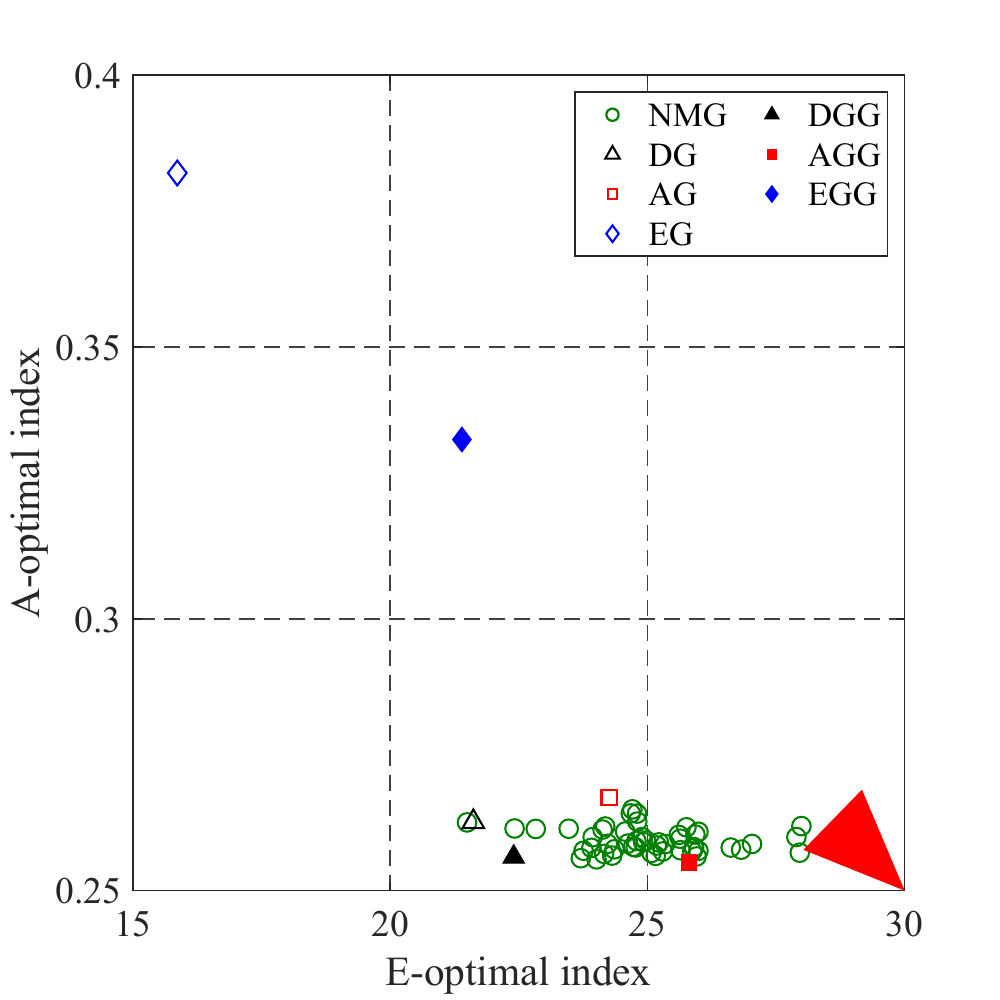}
    \caption {Relationship between the indices of on A- and E-optimality for $p=20$, $n=1000$ and $L_{\rm{\max}}=50$.}
    \label{fig:ParetoAE}
\end{figure}

Figs.~\ref{fig:ParetoDA}, \ref{fig:ParetoDE}, and \ref{fig:ParetoAE} show the Pareto-optimal solutions obtained by the NMG method. In Figs.~\ref{fig:ParetoDA}, \ref{fig:ParetoDE}, and \ref{fig:ParetoAE}, the values of optimal indices are plotted in the objective-function space: the D-optimal index against the A-optimal index, the D-optimal index against the E-optimal index, and the A-optimal index against the E-optimal index, respectively. The red triangles denote optimization directions in the objective-function space. The results of sensor selection by the pure greedy methods and the group greedy methods are also plotted together for comparison.
The number of sensor sets $L_{\rm{\max}}$ is set to 50 in the NMG and group greedy methods, and the number of sensors selected is set to 20 in all methods. 

Figs.~\ref{fig:ParetoDA}, \ref{fig:ParetoDE}, and \ref{fig:ParetoAE} show that the NMG method is able to find solutions superior to those found by the EG, EGG, AG, and AGG methods for all the optimality indices. The solutions obtained by the DG and DGG methods appear around the Pareto-optimal solutions obtained by the NMG method. Furthermore, the index of D-optimality of the solution obtained by the DGG method is higher than that for the best solution obtained by the NMG method in terms of the D-optimality, as discussed in the previous section.

Table~\ref{table:correlation} shows the correlation coefficients for each pair of the three optimality criteria. The results are analyzed using interactive Scatter Plot Matrix (iSPM), which is an optimal-solution visualization tool\cite{tatsukawa2017ispm}. 
It should be noted that the correlation coefficients in the case of $L_{\rm{\max}}=20$ and 50 are focused on in this subsection, since the correlation coefficients in the case of $L_{\rm{\max}}=5$ and 10 are not considered reliable due to the limited number of Pareto-optimal solutions ($L_{\rm{\max}}$). 

The magnitude of the correlation coefficient for the indices of D- and E-optimality is large and the sign is negative for the cases in which $L_{\rm{\max}}=20$ and 50. Thus, the D-optimality appears to be strongly related to the E-optimality, and there is a trade-off between the indices of D- and E-optimality. As shown in Fig.~\ref{fig:ParetoDE}, the shape of the Pareto-optimal front reveals the trade-off in the objective-function space of D- and E-optimality. When performance increases in terms of the D-optimality, performance decreases in terms of the E-optimality. 
On the other hand, the magnitude of the correlation coefficient for the indices of A- and E-optimality is relatively small in the case of $L_{\rm{\max}}=20$ and 50. Fig.~\ref{fig:ParetoAE} shows that no trade-off between the indices of A- and E-optimality. This indicates that the correlation between the indices of A- and E-optimality is weak, and the optimization using the A-optimality-based greedy methods is expected not to degrade performance in terms of the E-optimality compared to those based on D-optimality. This supports the results discussed in \cite{nakai2021effect}.
The indices of E-optimality obtained by the DG and AG methods were reported to be higher than that obtained by the EG method due to the lack of submodularity of the objective function based on E-optimality \eqref{eq:obj_eig}, and the AG method is superior to the DG method in terms of E-optimality. 
In addition, Table ~\ref{table:correlation} shows that the magnitude of the correlation coefficient for D- and A-optimality is small, and its sign varies depending on $L_{\rm{\max}}$ in the case of $L_{\rm{\max}}=20$ and 50. Fig.~\ref{fig:ParetoDA} shows no apparent trade-off in the objective-function space. Thus, the correlation between the indices of D- and A-optimality appears to be weak, and there is no trade-off between the two indices. 
% \begin{figure*}[htbp]
%     \centering
%     \includegraphics[width=6in]{figs_pareto/fig_iSPMlayout_2.pdf}
%     \caption{Relationship of three objective functions based on the optimal design. The figure is plotted using iSPM as a reference.}
%     \label{fig:Pareto}
% \end{figure*}

\begin{center}
\begin{table}[!tbp]
% \begin{threeparttable}[ht]
    \small
    \centering
    \caption{Correlation coefficient of the D-, A-, and E-optimal indices for $p=20$.} 
    \label{table:correlation}
    \begin{tabular}[t]{ccc}
    \hline
    Optimality & $L_{\rm{max}}=20$ & $L_{\rm{max}}=50$\\
    \hline
    D-A & 0.28 & -0.19 \\
    D-E & -0.84 & -0.88 \\
    A-E & 0.40 & 0.19 \\
    \hline
    \end{tabular}
% \end{threeparttable}
\end{table}
\end{center}

\section{Conclusions}
\label{sec:con}
We have proposed an NMG method for multiple set functions and applied it to the multi-objective optimization problem of the sensor selection based on the optimal experimental designs. \color{\ReviewerA}We theoretically showed that the NMG method falls into the class of the relaxed greedy methods, and that the relaxed greedy methods have performance guarantee when employing the monotone submodular function as well as a pure greedy method. Then, we applied the proposed method to the sensor selection problem. \color{black} The proposed method determines the sets of sensors based on nondominated solutions by simultaneously evaluating multi-objective functions based on D-, A-, and E-optimality. The method was applied to a randomly generated dataset and evaluated by comparing the performance of the method with those of various pure greedy and group greedy methods.

The results showed that the proposed method was able to outperform the pure greedy methods for all indices of the D-, A-, and E-optimality. It also showed the same or better performance when compared with the group greedy methods for indices of the three optimality, except in the case of the index of D-optimality obtained by the D-optimality-based group greedy method in the oversampling situation. The superior performance of the proposed method is mainly attributable to the increase in the search space and the introduction of Pareto dominance in multi-objective optimization.
On the other hand, the computational time for the proposed method is greater than that for the pure and group greedy methods primarily because of the nondominated sorting using the ENS-SS algorithm and the evaluation of all indices of the three optimality. However, the time complexity of the nondominated sorting can be effectively reduced by terminating the ENS-SS based on the number of reserved sets of sensors, and the resulting overhead of the ENS-SS is reasonable. 

The relationship between the three optimality was discussed based on the Pareto-optimal solutions obtained by the proposed method. It was found that there is a trade-off between the indices of D- and E-optimality, and that there is no correlation between the indices of A- and E-optimality. Therefore, the A-optimality-based greedy method is more suitable than the D-optimality-based greedy method for maximization of the index of E-optimality, where the objective function lacks submodularity, as previously reported.

The proposed method was shown to successfully overcome the difficulties of the previously-proposed greedy methods, and achieve an improved level of performance in terms of the standard indices of optimal experimental design.
In a practical sensor selection problem, it is also important to minimize the cost of arranging the sensors. Although it is difficult to effectively incorporate cost when defining the objective function in the typical greedy methods (e.g., using a hyperparameter to adjust the weights of the cost and index of optimality), the proposed method allows the user to easily introduce costs into the formulation.

Finally, it should be noted that the sensor optimization method can also be applied for actuator optimization problems through the dual problem relationship between the sensor and actuator selection problems. In addition, the idea proposed in the present study can improve the performance of various optimizations using greedy methods, not just sensor/actuator optimization. 

\section*{Acknowledgment}
This work was supported by JST CREST (JPMJCR1763), ACT-X (JPMJAX20AD), FOREST (JPMJFR202C), Moonshot R\&D – MILLENNIA Program (JPMJMS2287) and JSPS KAKENHI(21K14070), Japan.
% T.~Nonomura and K.~Yamada were supported by Japan Science and Technology Agency [Moonshot R\&D – MILLENNIA Program] Grant Number JPMJMS2287.

% \appendices
% \section{Efficient implementations in Matlab}
% The computation of the each objectives are conducted intensively-repeatedly for so enormous number of candidates of sensor. This section elaborates the actual implementations for MATLAB\limited of the computations that fully enjoy the 1-rank nature of the A- and D-optimality criteria.
% 

% \clearpage

\bibliographystyle{IEEEtran}
\bibliography{xaerolab}

\end{document}